\documentclass[10pt,final,doublecolumn]{IEEEtran}
\IEEEoverridecommandlockouts
\usepackage{amsmath}
\usepackage{amssymb}
\usepackage{amsthm}
\usepackage{mathrsfs}
\usepackage{latexsym}
\usepackage{graphicx}
\usepackage{bbding}
\usepackage{indentfirst}
\usepackage{cases}
\usepackage{supertabular}
\usepackage{algorithm,algorithmic}
\usepackage{subeqnarray}
\usepackage{color}
\usepackage{bm}
\usepackage{stfloats}

\usepackage{subfigure}
\usepackage{array}
\usepackage{stfloats}
\usepackage{algorithmic}
\usepackage{float}
\usepackage{algorithm}
\allowdisplaybreaks[4]

\begin{document}
\title{On the Performance of Integrated Satellite-Terrestrial Maritime Communications}
\author{\IEEEauthorblockN{
Kaiwei Xiong, Xiaoming Chen, and Ming Ying}
\thanks{Kaiwei Xiong, Xiaoming Chen, and Ming Ying are with the College of Information Science and Electronic Engineering, Zhejiang University, Hangzhou 310027, China (e-mail:\{xiong\_kaiwei, chen\_xiaoming, and ming\_ying\}@zju.edu.cn).}
}\maketitle

\begin{abstract}
In this paper, we present an integrated terrestrial and satellite maritime communication system, where a shore-based terrestrial base station (TBS) and a low Earth orbit (LEO) satellite cooperatively provide wide-area communication services to maritime users. We conduct performance analysis for the integrated satellite-terrestrial maritime communication system. Specifically, we analyze the transmission rate and coverage probability of near-shore and off-shore users respectively according to the maritime communication environment. Besides, in order to better understand the impact of some key parameters, we also make asymptotic analysis in some special cases. Further, we design an optimization algorithm to maximize the coverage probability of near-shore users by adjusting the transmission power of TBS and the LEO satellite, while ensuring the both off-shore and near-shore users can meet the minimum communication rate requirements. Finally, extensive numerical analysis results verify the accuracy of the theoretical results and the effectiveness of the proposed optimization algorithm in the maritime communication system.	
\end{abstract}


\section{Introduction}
In recent years, maritime activities have experienced a significant increase \cite{Maritime transport1}, \cite{Maritime transport2}. Firstly, the recovery of global trade and the growth of e-commerce to drive the increase in the demand for goods transportation. Secondly, with the increasing demand for maritime resources \cite{growing demand1}-\cite{growing demand3}, some maritime fishery activities are also expanding, which is reflected in near-shore aquaculture and off-shore fisheries. The increasing number of maritime activities makes the increase of maritime users and maritime data, which also puts forward higher demands and challenges for maritime communication. As the number of ships and users at sea continues to rise, the demand for safe, reliable, and rapid maritime communications becomes increasingly crucial. Besides, the need for real-time data exchange to transmit information is also important, such as real-time transmission of location, weather and ocean conditions, which is better for navigation planning and decision making. However, the current maritime communication is still far behind the terrestrial communication \cite{6G1}, \cite{6G2}, and the maritime communication still faces many challenges.

At present, maritime communication has been faced with many challenges, which affect the efficiency, security and reliability of communication. The area of the ocean is huge, and the traditional shore-based communication network cannot cover all the sea \cite{TBS1}-\cite{TBS service}. Especially when maritime users are far from the coastal area, the signal strength and stability will be greatly reduced. The ocean area is vast, and only shore-based stations network cannot cover all the sea areas. The only satellites network also cannot provide high throughput and timely information transmission. The development of the robust maritime communication infrastructure necessitates the integration of the LEO satellite and terrestrial systems, leveraging the extensive coverage capabilities of LEO satellite system alongside the high-throughput benefits of cellular systems. Such a hybrid approach is essential to address the growing demands and challenges faced by maritime communication systems \cite{hybrid1}-\cite{hybrid3}.

Furthermore, a large number of studies have been performed on how to construct and utilize the integrated satellite-terrestrial communication systems more efficiently \cite{R1}-\cite{R2}, such as the secure communication, relay communication and massive access. In \cite{R3}, the authors proposed a power control algorithm for the integrated satellite-terrestrial network, derived the average transmission power for theoretical performance evaluation, and revealed the influence of various key parameters on satellite communication. In \cite{R4}, the authors investigated the outage probability and corresponding asymptotic outage behavior of the integrated satellite-terrestrial network in the unified form over shadowed-Rician satellite links’ channels and Nakagami-m as well as Rician terrestrial links’ channels. Besides, regarding the reconfigurable intelligent surface (RIS) in assisting maritime wireless communication, some papers \cite{R5}-\cite{R6} introduced RIS to maximize the confidentiality rate of base stations while meeting the signal reception constraints, RIS's harvested power threshold, and the total transmission power budget.

The performance analysis of maritime communication network is an important link to ensure the efficiency and safety of maritime activities. Analyze how the combination of LEO satellite communications \cite{chen} and terrestrial base stations affects overall network performance, including signal coverage, data transmission efficiency, and cost. Maritime communication is different from the terrestrial communication systems, the number of maritime stations is small. Besides, the distribution of maritime users is greatly sparse. Thus, it is extremely important to evaluate the coverage area of network signals to ensure that maritime users can receive stable signals in different locations. In maritime communication network, many parameters affect the performance and reliability of the network. Signal strength is affected by transmission power, propagation distance, environmental interference and obstacles. Furthermore, the wave fluctuation and some obstacles will also cause some refraction and reflection of the signals, which will inevitably weaken the signal strength. Apart from these environmental impacts, interference between maritime users can not be avoided. The beam of the LEO satellite is often large, it may also cause interference to the near-shore users communicating with the TBS \cite{wq}. Thus it is also necessary to balance the relationship between signal strength and interference by adjusting the transmit power.

Stochastic geometry is a useful mathematical tool that combines stochastic processes and geometry, which is widely used in performance analysis, network design, and other fields to provide analytical results for key performance metrics. Stochastic geometry provides a powerful tool for understanding and optimizing complex systems due to that it can analyze easily the various types of wireless networks. Therefore, based on the stochastic geometry, we characterize the integrated satellite-terrestrial maritime network, and explain the influence of key parameters on signal transmission.

Stochastic geometry has been widely used in satellite communication networks and terrestrial base station networks. For wireless communication networks with downlink terrestrial base stations, \cite{1} provides a comprehensive framework for analyzing data rate and coverage of single layer base station deployments, in which these expressions generate useful insights for communication system design. In \cite{3}, the analysis of downlink is extended to the uplink, and the analytical expression of spectral efficiency when the base stations and antennas increase gradually is derived by analyzing the performance of wireless networks with interference. In \cite{4}, the hybrid terrestrial network is structured as a superposition of multi-layer base stations with different distribution, fading parameters, path loss index, density and transmit power. By using stochastic geometry to model and analyze the interference in other cells, an analytical framework is proposed to derive the transmission data rates of hybrid terrestrial communication systems.

Using stochastic geometry can not only analyze the performance of the terrestrial cellular systems, but also have many related works to analyze the satellite communication network. In \cite{6}, the stochastic geometry method is used to model the LEO satellite constellation, providing low complexity end-to-end availability analysis and link coverage performance estimation. In \cite{5}, the mathematical formulations for key performance metrics, including network coverage probability and mean transmission rate in downlink scenarios, are established for general LEO satellite constellations. In \cite{7}, considering the impact of atmospheric attenuation and beam coverage angle on the overall coverage probability, a new interference simplification method is proposed. In \cite{8}, it introduces a finer-grained analysis of LEO satellite systems modeled by homogeneous Poisson point processes (HPPP), and investigates the moments and distribution of conditional coverage probabilities based on the specified point processes.

Although these previous works have provided valuable insights into the performance metrics of single layer terrestrial cellular networks and LEO satellite networks. Besides, these works are mostly centered around the situation on the ground, and there are still limitations in the study of maritime communication networks. The influence of parameters on the performance of maritime communication network is still unclear. The traditional closed-form expressions of system performance based on Rayleigh fading can not directly be extended to Rician fading, because its probability density function contains the complex Bessel function. In order to address the limitations of performance analysis studies and more fully understand coverage performance, we focused on the impact of modeling actual maritime communication channels. Specifically, we use the two-ray path loss model and Rician fading as small-scale fading to established a channel model, which is closer to the real maritime communication environment. Various modeling works and channel measurements have been conducted to analyze the impact of maritime environments and system parameters on maritime communications, where the Rician fading and two-ray channel models have been verified and widely applied in the maritime communication \cite{YZ1}-\cite{YZ6}. We propose an approximation method that can transform the Rician distribution into a finite Dirichlet series, which is well for the convenience of subsequent derivation and calculation. Furthermore, we propose an analytical method to simulate the performance of the downlink. The main contributions of this paper are outlined as follows.

\begin{enumerate}
	
	\item We provide an integrated satellite-terrestrial maritime communication framework, which leverages TBS to deliver connectivity for near-shore users while utilizing LEO satellite to ensure network accessibility for off-shore users. In such a framework, the wide-area coverage can be achieved for the most sparse maritime users.
	
	\item We conduct performance analysis for the integrated satellite-terrestrial maritime communication system. Specifically, we derive closed-form expressions of the transmission rate and coverage probability and for near-shore and off-shore users, respectively. We further make the asymptotic analysis for some key parameters, and reveal their impacts on the performance metrics.
	
	\item We propose an optimization algorithm to maximize the coverage probability of near-shore users by adjusting the transmit power of LEO satellite and TBS while guaranteeing the maritime users' data rates.
	
\end{enumerate}

The rest of this article is organized as follows. Section \uppercase\expandafter{\romannumeral2} introduces the geometric model and system model. In section \uppercase\expandafter{\romannumeral3}, we derive the tractable forms of the average rate and coverage probability for off-shore and near-shore users, respectively. In section \uppercase\expandafter{\romannumeral4}, the transmit power of satellite and TBS is jointly designed to maximize the coverage probability of near-shore users while satisfying QoS requirements of both near-shore and off-shore users. Section \uppercase\expandafter{\romannumeral5} provides simulation results to verify the accuracy of derived formula and the effectiveness of proposed algorithm. To conclude, section \uppercase\expandafter{\romannumeral6} summarizes this paper.

\emph{Notations}: This paper represents matrices and vectors by bold letters, $\|\cdot\|$, $(\cdot)^H$, $|\cdot|$ to denote $L_2$-norm of a vector, conjugate transpose and absolute value, respectively. $Q(\cdot)$ and $J_0(\cdot)$ are employed to represent the Marcum Q-function and the zeroth order of the first-kind of Bessel function, respectively.

\section{System Model}
In this section, we introduced an integrated satellite-terrestrial maritime communication system. Besides, maritime communication systems usually work in the VHF band of 30 to 300MHz. Thus, we consider that the LEO satellite and TBS work over the same spectrum in the VHF band. In what follows, we present the geometry model and channel model of the considered maritime communication system.

\subsection{Geometry Model}
\begin{figure}
		\centering
		\includegraphics [width=0.52\textwidth] {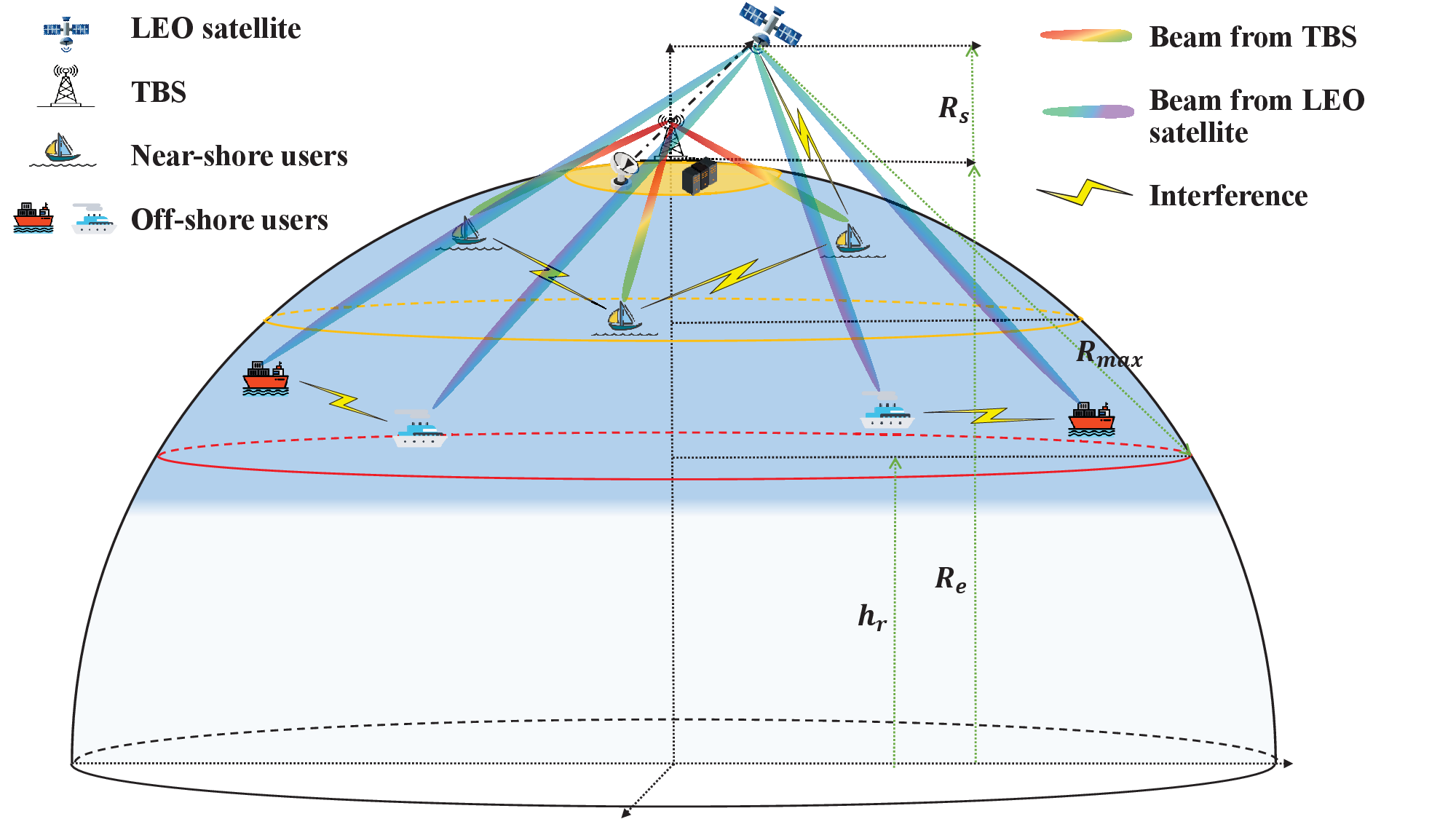}
		\caption {Geometry model of the integrated satellite-terrestrial maritime communication system.}
		\label{system}
\end{figure}

As shown in Fig. $\ref{system}$, we propose a hybrid maritime communication system, within which TBS and LEO satellite collaboratively establish a comprehensive connectivity framework for diverse maritime users. Moreover,  $R_s$ and $R_e$ denote the height of the LEO satellite and the radius of the Earth, respectively. Without loss of generality, the TBS is deployed to provide services for near-shore users, and the LEO satellite is operated in the orbit with a radius of $R_e+R_s$ providing services for off-shore users. According to Archimedes's Hat-Box Theorem \cite{cap}, the coverage range of the LEO satellite forms a spherical cap on Earth, which can be computed as
	\begin{equation}\label{cap}
	\begin{aligned}
		|\mathcal{A}|=2\pi R_e(R_e+R_s).
	\end{aligned}
\end{equation}

Furthermore, the spherical cap $\mathcal{A}_r$ which contains the users with the distance of less than $r$ from LEO satellite
\begin{equation}
	\begin{aligned}\label{capr}
		|\mathcal{A}_{r}|=2\pi (R_e-h_r)(R_e+R_s),
	\end{aligned}
\end{equation}
where $h_r$ can be expressed in terms of $r$ as
\begin{equation}\label{hr}
	\begin{aligned}
		h_r=\frac{R_s^2+2R_eR_s-r^2}{2R_e},
	\end{aligned}
\end{equation}
where $r$ denotes the distance from the maritime user to the LEO satellite. Similarly, the coverage range of the TBS forms a circle area with radius $R_0$, the area of which is given by
\begin{equation}
	\begin{aligned}\label{capt}
		|\mathcal{A}_{t}|=\pi R_0^2.
	\end{aligned}
\end{equation}

The near-shore and off-shore users obeys the HPPP with density $\lambda_1$ and $\lambda_2$, respectively. Specifically, the number of near-shore users $M_1$ and off-shore users $M_2$ are modeled by a Poisson distributed that have a mean value of  $|\mathcal{A}_{t}|$ and $|\mathcal{A}_{r}|$, respectively, which can be expressed as
\begin{equation}\label{M1}
	\begin{aligned}
		\mathbb{P}(M_1=m)=\text{exp}(- |\mathcal{A}_{t}|\lambda_1) \frac{(|\mathcal{A}_{t}| \lambda_1)^m}{m!},
	\end{aligned}
\end{equation}

\begin{equation}\label{M2}
	\begin{aligned}
		\mathbb{P}(M_2=n)=\text{exp}(-|\mathcal{A}_{r}| \lambda_2) \frac{(|\mathcal{A}_{r}| \lambda_2)^n}{n!}.
	\end{aligned}
\end{equation}

\subsection{Channel Model}
In general, the maritime communication channel from the TBS to the user is position independent because of the sparsity of scatters and the absence of significant obstacles. Compared with the terrestrial environment, due to the evaporation of a large amount of seawater, the distribution of the atmosphere on the sea surface is uneven. Shore-ship communication and ship-to-ship communication are not only susceptible to sea surface conditions, but also to atmospheric conditions such as temperature, humidity and wind speed. Moreover, the height of the antennas and angle of maritime wireless channel vary greatly with wave fluctuations. The fading channel is particularly sensitive to these parameters, which may cause frequent link interruptions. However, some robust beamforming algorithms can be used to resist the impacts of the maritime environment on communication.

Therefore, the path loss of maritime channel is primarily determined by the line-of-sight (LoS) and reflected path over the ocean surface, typically modeled using the two-ray channel model \cite{two ray}. For the two-ray channel in maritime communications, the signal from the clear LoS path is the predominant signal while the signal over the reflected path has a less impact on the overall signal strength. Hence, the Rician fading channel model is employed to characterize the maritime communication channel. Mathematically, the channel coefficient between the near-shore user $k$ and TBS can be modeled as\footnote{Due to the deterministic LEO's trajectory, the Doppler frequency shift caused by the high mobility of the LEO satellite can be estimated and compensated in advance \cite{DP1}-\cite{DP2} by maximum likelihood Doppler estimation, Kalman filter and geometric analysis. Thus, we assume that the effect of the Doppler shift can be mitigated.}
\begin{equation}\label{channel1}
	\begin{aligned}
		H_{1,k}=\bigg[\frac{\zeta \sqrt{G_1P_1}}{2\pi d_k}\sin(\frac{2\pi H_tH_r}{\lambda d_k})\bigg]h(K_1 ),
	\end{aligned}
\end{equation}
where $\zeta$ denotes the carrier wavelength. $G_1$ and $P_1$ denote the transmit antenna gain and the transmit power of TBS, respectively. $d_k$ denotes the distance from TBS to the near-shore user $k$, $H_t$ and $H_r$ are the effective antenna height of TBS and users, respectively. Moreover, $h(K_1)$ represents the Rician fading, which is modeled as \cite{ym2}
\begin{equation}\label{Rician}
	\begin{aligned}
		h(K_1)=\sqrt{\frac{1}{1+K_1}}h_1^{\text{NLoS}}+\sqrt{\frac{K_1}{1+K_1}}h_1^{\text{LoS}},
	\end{aligned}
\end{equation}
where $K_1$ denotes the Rician factor of near-shore area. $h_1^{\text{NLoS}}$ denotes the non-line of sight (NLoS) components of near-shore area, which is an identical and independent distributed random variable that follows standard complex Gaussian distribution, and $h_1^{\text{LoS}}$ denotes the LoS component of near-shore area. Then, the probability density function (PDF) of the Rician fading model can be expressed as \cite{Q}
\begin{equation}\label{PDF}
	\begin{aligned}
		f_{h(K_1)}(x)=\frac{x}{\rho^2}\text{exp}\bigg(\frac{-(x^2+v^2)}{2\rho^2}\bigg)J_0(\frac{xv}{\rho^2}),
	\end{aligned}
\end{equation}
where $J_0(\cdot)$ denotes the zeroth order of the first-kind of Bessel function, $\rho=\sqrt{\frac{1}{2+2K_1}}$ is the power of multipath signal component, and $v=\sqrt{\frac{K_1}{1+K_1}}$ represents the peak of main signal amplitude. Therefore, the cumulative distribution function (CDF) of Rician channel is derived as
\begin{equation}\label{CDF}
	\begin{aligned}
		F_{h(K_1)}(x)=1-Q_1\bigg(\frac{v}{\rho},\frac{x}{\rho}\bigg),
	\end{aligned}
\end{equation}
where $Q_1(a,b)$ is the Marcum Q-function defined as $Q_1(a,b)=\int_{b}^{\infty}x\text{exp}(-\frac{x^2+a^2}{2})J_0(ax)dx$.

Taking into account the signal transmission properties between the off-shore users and LEO satellite \cite{channel}-\cite{wq2}, the space-to-ocean channel between the off-shore user $m$ and LEO satellite can be mathematically characterized as as
\begin{equation}\label{channel2}
	\begin{aligned}
		H_{2,m}=g_r\sqrt{r_{\text{FSL}}P_2 G_2}h(K_2),
	\end{aligned}
\end{equation}
where $P_2$ and $G_2$ denote the transmit power and the transmit antenna gain of the LEO satellite, respectively. $h(K_2)$ denotes the small-scale fading of space-to-sea channel, which can be modeled as
\begin{equation}\label{Rician}
	\begin{aligned}
		h(K_2)=\sqrt{\frac{1}{1+K_2}}h_2^{\text{NLoS}}+\sqrt{\frac{K_2}{1+K_2}}h_2^{\text{LoS}},
	\end{aligned}
\end{equation}
where $K_2$ represents the Rician factor of off-shore area. $h_2^{\text{NLoS}}$ denotes the NLoS components of off-shore area, and $h_2^{\text{LoS}}$ denotes the LoS component of off-shore area. Furthermore, $r_{\text{FSL}}$ represents the free space loss, which is given by
\begin{equation}\label{FSL}
	\begin{aligned}
		\ r_{\text{FSL}}=\frac{\zeta^{2}}{(4\pi D_m)^{2}},
	\end{aligned}
\end{equation}
where $D_m$ is the distance between the $m$ off-shore user and  LEO satellite. Moreover, $g_r$ is the rain attenuation coefficient.

In this section, we introduce the hybrid maritime communication system, including geometry model and channel model. In what follows, we will analyze the performance of the integrated satellite-terrestrial maritime communication system based on the above geometry and channel models.

\section{Performance Analysis}
In this section, we concentrate on the analysis of the performance of the considered integrated satellite-terrestrial martime communication system in terms of the average rate and coverage probability. Subsequently, we analyze the impact of key system parameters on the overall performance.

\subsection{Distance Distribution}
With the aim of calculating the performance, the visibility probability of TBS and LEO satellite is essential. Therefore, we first derive the visibility probabilities for both the LEO satellite and TBS. Specifically, the LEO satellite visible probability can be expressed as
\begin{equation}
	\begin{aligned}\label{Psatvis}
		P_{\mathrm{Sat}}^{\mathrm{vis}}&=\mathbb{P}\big(\Phi(\mathcal{A})>0\big) \\
		&=1-\exp(-\lambda_2|\mathcal{A}|),
	\end{aligned}
\end{equation}
where $\Phi(\mathcal{A})>0$ represents the number of off-shore users in the coverage of LEO satellite. Similarly, the visible probability of TBS is given by
\begin{equation}\label{PTBSvis}
	\begin{aligned}
		P_{\mathrm{TBS}}^{\mathrm{vis}}&=\mathbb{P}\big(\Phi(\mathcal{A}_t)>0 \big)\\
		&=1-\exp(\lambda_1|\mathcal{A}_t|),
	\end{aligned}
\end{equation}
where $\Phi(\mathcal{A}_t)>0$ represents the number of near-shore users in the coverage of TBS. Next, we compute the probability that the smallest transmission distance between the off-shore users and LEO satellite is larger than $r$, conditioned that on the presence of more than one off-shore user exists in region $\mathcal{A}$. We determine the conditional probability of no user occurrence with region $\mathcal{A}_r$, expressed as $\Phi(\mathcal{A}_r)=0$. Considering $\Phi(\mathcal{A})>0$, we have
\begin{equation}\label{Pr}
	\begin{aligned}
		&P[D_m>r|\Phi(\mathcal{A})>0] \\
		&= \mathbb{P}[\Phi(\mathcal{A}_r)=0|\Phi(\mathcal{A})>0] \\
		&=\frac{\mathbb{P}[\Phi(\mathcal{A}_r)=0,\Phi(\mathcal{A})>0]}{\mathbb{P}[\Phi(\mathcal{A})>0]}\\
		&\overset{(a)}{=}\frac{\mathbb{P}[\Phi(\mathcal{A}/\mathcal{A}_r)>0] \times \mathbb{P}[\Phi(\mathcal{A}_r)=0]}{\mathbb{P}[\Phi(\mathcal{A})>0]}\\
		&=\frac{\big(1-\mathbb{P}[\Phi(\mathcal{A}/\mathcal{A}_r)=0]\big)\times \mathbb{P}[\Phi(\mathcal{A}_r)=0]}{1-\mathbb{P}[\Phi(\mathcal{A})=0]}\\
		&\overset{(b)}{=}\frac{e^{-\lambda_2 |\mathcal{A}_r|}\times\big(1-e^{-\lambda_2|\mathcal{A}/\mathcal{A}_r|}\big)}{1-e^{-\lambda_2|\mathcal{A}|}}\\
		&=\frac{e^{-\lambda_2 |\mathcal{A}_r|}-e^{-\lambda_2\mathcal{A}}}{1-e^{-\lambda_2|\mathcal{A}|}},
	\end{aligned}
\end{equation}
where (a) is derived by the independence of PPP for $\mathcal{A}/\mathcal{A}_r$ and $\mathcal{A}_r$, and (b) comes from the LEO satellite visible probability. Plugging (\ref{Psatvis}) into (\ref{Pr}), we can obtain the conditional CCDF as
\begin{equation}
	\begin{aligned}\label{rCCDFsat}
		&F_{r_{\mathrm{Sat}|\Phi(\mathcal{A})>0}}(r)\\
		&=\frac{\exp(-2\lambda_2 \pi R_e(R_e+R_s))}{1-\exp(-\lambda_2 \pi R_e(R_e+R_s))}\\
		&*\frac{\big[\exp(-\lambda_2 \pi \frac{(R_e+R_s)(r^2-R_e(R_s+R_e))}{R_e})-1\big]}{1-\exp(-\lambda_2 \pi R_e(R_e+R_s))}.
	\end{aligned}
\end{equation}

With (\ref{rCCDFsat}), we can obtain the PDF of conditional distribution with the smallest maritime off-shore user transmission distance as
\begin{equation}\label{rPDFsat}
	\begin{aligned}
		&f_{r_{\mathrm{Sat}}|\Phi(\mathcal{A})>0}(r)\\
		&=2\lambda_2 \pi r*\frac{R_e+R_s}{R_e}*\frac{\exp(\lambda_2 \pi (R_s+R_e)^2)}{\exp(2\lambda_2 \pi R_e(R_s+R_e)-1}\\
		&*\exp\bigg(-\lambda_2 \pi \frac{R_s+R_e}{R_e}r^2\bigg),
	\end{aligned}
\end{equation}
where $R_{\min}<r<R_{\max}$ with $R_{\min}=R_s$, and $R_{\max}=\sqrt{R_s(R_s+2R_e)}$.

Similarly, we can compute the distribution fo the nearest maritime near-shore users transmission distance. Based on the user distribution model, the CCDF of distance between the nearest near-shore user and TBS is modeled as
\begin{equation}
	\begin{aligned}\label{rCCDFTBS}
		&F_{r_{\mathrm{TBS}|\Phi(\mathcal{A}_t)>0}}(r)=\exp\big(-\lambda_1 \pi (r^2-R_0^2)\big).
	\end{aligned}
\end{equation}

Accordingly, the PDF of conditional distribution of the nearest maritime near-shore user transmission distance can be given by
\begin{equation}\label{rPDFTBS}
	\begin{aligned}
		&f_{r_{\mathrm{TBS}|\Phi(\mathcal{A}_t)>0}}(r)=2\lambda_1 \pi r \exp\big(-\lambda_1 \pi (r^2-R_0^2)\big),
	\end{aligned}
\end{equation}
where $r<R_0$.

\subsection{Coverage Probability}
In this subsection, we compute the coverage probabilities for the LEO satellite and TBS. Firstly, we give the expressions for signal-to-interference-plus-noise ratio (SINR), which are given by
\begin{equation}\label{SINR1}
	\begin{aligned}
		\Upsilon_{1,k}=\frac{\bigg[\frac{\zeta \sqrt{G_1P_1} }{2\pi d_k^2}|h(K_1)|\bigg]^2}{I_1+I_2+\sigma_{1}^2},
	\end{aligned}
\end{equation}

\begin{equation}\label{SINR2}
	\begin{aligned}
		\Upsilon_{2,m}=\frac{\bigg[\frac{g_r \zeta}{4\pi D_m}\sqrt{P_2 G_2} |h(K_2)|\bigg]^2}{I_2+\sigma_{2}^2},
	\end{aligned}
\end{equation}
where $\sigma_{1}^{2}$ and $\sigma_{2}^2$ are the variance of additive white Gaussian noise (AWGN). $I_1$ and $I_2$ are the aggregated interference power from the TBS and LEO satellite, respectively, which can be expressed as
\begin{equation}\label{I1}
	\begin{aligned}
		I_1=\sum_{i=1,i\ne k}^{M_1} \bigg[\frac{\zeta \sqrt{G_1P_1} }{2\pi d_i^2}|h(K_1)|\bigg]^2 ,
	\end{aligned}
\end{equation}

\begin{equation}\label{I2}
	\begin{aligned}
		I_2=\sum_{n=1, n\ne m}^{M_2} \bigg[\frac{g_r \zeta}{4\pi D_n}\sqrt{P_2 G_2}|h(K_2)|\bigg]^2 ,
	\end{aligned}
\end{equation}
where $M_1$ and $M_2$ denote the number of interfering neighbor users of TBS's coverage area and LEO satellite's area, respectively. The coverage probability is define as the probability of received SINR at the maritime users greater than a predefined threshold, which is given by
\begin{equation}
	\begin{aligned}
		P^{\mathrm{cov}} \triangleq \mathbb{P}(\Upsilon > \gamma),
	\end{aligned}
\end{equation}
where $\gamma$ denotes the threshold SINR for reliable communication. According to (24), the coverage probability of near-shore and off-shore users are defined as
\begin{equation}
	\begin{aligned}
		\mathcal{P}_{n} = P^{\mathrm{vis}}_{\mathrm{TBS}}*P^{\mathrm{cov}}_1,
	\end{aligned}
\end{equation}

\begin{equation}
	\begin{aligned}
		\mathcal{P}_{o} = P^{\mathrm{vis}}_{\mathrm{Sat}}*P^{\mathrm{cov}}_2,
	\end{aligned}
\end{equation}
where $P^{\mathrm{vis}}_{\mathrm{TBS}}$ and $P^{\mathrm{vis}}_{\mathrm{Sat}}$ are the visible probabilities of TBS and LEO satellite, respectively. The conditional coverage probability $P^{\mathrm{cov}}_1$ and $P^{\mathrm{cov}}_2$ characterize the SINR distribution observed at a reference receiver under the presence of either LEO satellite or TBS within the spherical cap region $\mathcal{A}$. Subsequently, we analyze the coverage probability given the existence of at least one near-shore user in $\mathcal{A}$. Through conditional probability analysis based on $\Phi(\mathcal{A})>0$, the coverage probability conditioned on at least one near-shore user in $\mathcal{A}$ is expressed as
\begin{equation}\label{P1cov}
	\begin{aligned}
		P^{\text{cov}}_1=\mathbb{P}\big(\Upsilon_{1,k}>\gamma_1|\Phi(\mathcal{A})>0\big).
	\end{aligned}
\end{equation}

Further, we can derive the average of coverage probabilities over all possible positions of near-shore users in $\mathcal{A}$, which can be computed as
\begin{equation}\label{P1cov2}
	\begin{aligned}
		&P^{\mathrm{cov}}_1\\
		&=\mathbb{E}[\mathbb{P}(\Upsilon_{1,k}>\gamma_1|\Phi(\mathcal{A})>0)]\\
		&=\int_{R_{\min}}^{R_{\max}} \mathbb{E}\bigg(\mathbb{P}(|h(K_1)|^2>\frac{4 \pi^2 \gamma_1(I_1+I_2+\sigma_{1}^2)r^{-4}}{\zeta G_1P_1})\bigg)\\
		&\times f_{r_{\mathrm{TBS}|\Phi(\mathcal{A}_t)>0}}(r)dr.
	\end{aligned}
\end{equation}

However, it is difficult to calculate the expectation in (\ref{P1cov2}) since $h(K_1)$ follows the Rician distribution. To the end, we use the finite Dirichlet series to approximate the CDF of $h(K_1)$, which can be expressed as

\begin{equation}\label{jinsi}
	\begin{aligned}
     F(x) \approx 1-\sum_{n=1}^{N}a_n \exp(-b_n x) = F^{'}(x),
	\end{aligned}
\end{equation}
where $N$ is the general Dirichlet series terms, $a_n$ and $b_n$ are parameters, such that $b_n>0$ and $\sum_{n=1}^{N}a_n=1$. To evaluate the approximation error, we define the approximation error function $\epsilon(K)$ as
\begin{equation}\label{wucha1}
	\begin{aligned}
		\epsilon(K)=\int_{0}^{\infty}\bigg(F(x)-F^{'}(x)\bigg)^2dx.
	\end{aligned}
\end{equation}

Since the integration is still difficult to calculate, we expand it into a series form according to the definition of the integration. Specifically, the series form can be expressed as
\begin{equation}\label{wucha2}
	\begin{aligned}
		\hat{\epsilon}(K)=\delta \sum_{i=0}^{\infty}\bigg( F(i\delta)-F^{'}(i\delta)  \bigg)^2,
	\end{aligned}
\end{equation}
where $\delta$ is a small number and $\hat{\epsilon}(K) \rightarrow \epsilon(K)$ as $\delta \rightarrow 0$. Therefore, the required task is to find proper parameters $a_n$ and $b_n$ so that (\ref{wucha2}) is minimized. Thus, we can get the following optimization problem as
\begin{subequations}
	\begin{eqnarray}
		\mathcal{Q}:\underset{a_n,b_n}{\mathop{\min}}\,\!\!\!\!&&\!\! \delta \sum_{i=0}^{\infty}\bigg( F(i\delta)-F^{'}(i\delta) \bigg)^2 \label{OP1obj}\\
		\textrm{s.t.}\!\!\!\!&&\!\!  \sum_{n=1}^{N}a_n=1\\ \label{OP1st1}
		\!\!\!\!&&\!\!  b_n>0,   \forall n \in \{1,2,\dots,N\}\label{OP1st2}
	\end{eqnarray}
\end{subequations}

\begin{figure}[h]
	\centering
	\includegraphics [width=0.45\textwidth] {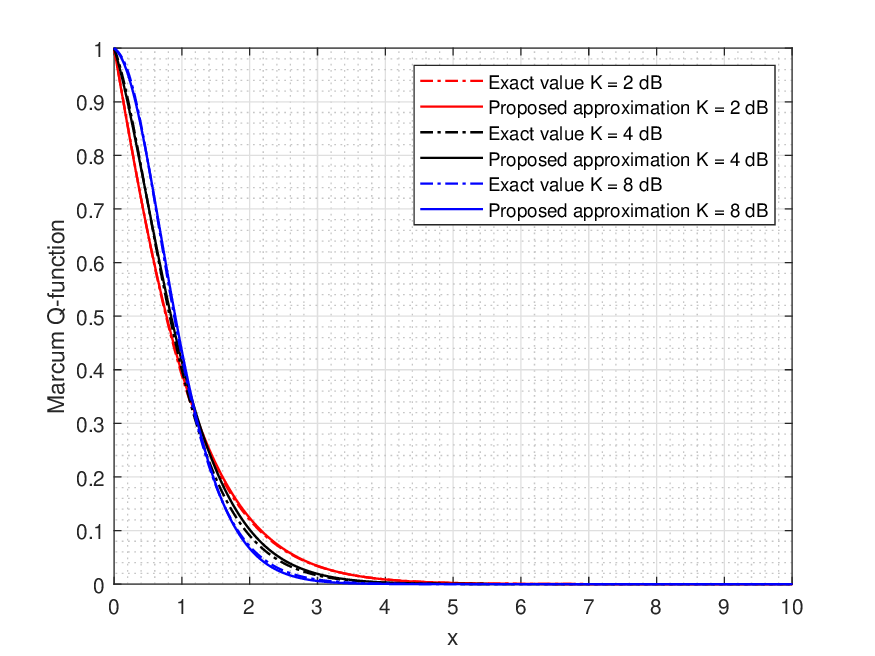}
	\caption {Comparison of the Marcum Q-function with the proposed approximation.}
	\label{Q}
\end{figure}

By using the genetic algorithm (GA) which is a classical intelligent method \cite{GA}, we can obtain the feasible values of $a_n^*$ and $b_n^*$ for (\ref{jinsi}) under the Rician factor $K=2.414$ dB. Specifically, Fig. \ref{Q} illustrates the comparison of the proposed approximation with the Marcum Q-function under various Rician factors. The overall level of accuracy of (\ref{jinsi}) is in general quite acceptable. As a consequence, $P_1^{\mathrm{cov}}$ can be straightforwardly rewritten as follows
\begin{equation}\label{P1cov3}
	\begin{aligned}
		&P^{\mathrm{cov}}_1\\
		&=\mathbb{E}[\mathbb{P}(\Upsilon_{1,k}>\gamma_1|\Phi(\mathcal{A})>0)]\\
		&=\int_{R_{\min}}^{R_{\max}} \sum_{n=1}^{N}a_n^* e^{-\frac{b_n^*\gamma_1\sigma_1^2r^{-4}}{\zeta G_1P_1}} \mathcal{L}_{I_1}\mathcal{L}_{I_2} f_{r_{\mathrm{TBS}|\Phi(\mathcal{A}_t)>0}}(r)dr,
	\end{aligned}
\end{equation}
where $\mathcal{L}_{I_{\mathrm{TBS}}}$ and $\mathcal{L}_{I_{\mathrm{Sat}}}$ denote the conditional Laplace transform of the interference of other near-shore users and the LEO satellite. Specifically, $\mathcal{L}_{I_{\mathrm{TBS}}}$ can be computed as
\begin{equation}
	\begin{aligned}
		&\mathcal{L}_{I_1}(s_1)\\
		&=\mathbb{E}[e^{-s_1I_1}|\Phi(\mathcal{A})>0]\\
		&\overset{(a)}{=}\exp \bigg(-\lambda_1\int_{x \in \mathcal{A}} \big(1-\mathbb{E} \big[e^{-s_1G_1P_1h^2(K_1)x^{-4}} \big]   \big)  dx \bigg)\\
		&\overset{(b)}{=}\exp \bigg(-\lambda_1\int_{x \in \mathcal{A}} 1-\frac{1}{(1+\frac{s_1P_1G_1r^{-4}}{m_1})^{m_1}}  dx \bigg)\\
		&\overset{(c)}{=}\exp \bigg(-2\pi \lambda_1 \int_{r}^{R_0} \bigg[1-\frac{1}{(1+\frac{s_1P_1G_1r^{-4}}{m_1})^{m_1}}\bigg]x  dx \bigg)\\
		&\overset{(d)}{=}\exp \bigg(-\pi \lambda_1 \bigg(\frac{s_1P_1G_1}{m_1}\bigg)^{\frac{1}{2}}\int_{(\frac{s_1P_1G_1}{m_1})^{-\frac{1}{2}}r^2}^{(\frac{s_1P_1G_1}{m_1})^{-\frac{1}{2}}R_0^2}\\
		& 1-\frac{1}{(1+y^{-2})^{m_1}}  dy \bigg),
	\end{aligned}
\end{equation}
where $x$ is the near-shore users and $m_1=\frac{(K_1+1)^2}{2K_1+1}$ is  a given constant of $K_1$, (a) is based on the application of the probability generating functional (PGFL) properties for PPP, (b) originates from the independence between the point process and the small-scale fading, (c) comes from $\frac{\partial |\mathcal{A}|}{\partial r}=2\pi r $, and (d) is the variable substitution operation $y=(\frac{s_1P_1G_1}{m_1})^{-\frac{1}{2}}x^2$ and $dy=2x(\frac{s_1P_1G_1}{m_1})^{-\frac{1}{2}}dx$. Similarly, $\mathcal{L}_{I_{\mathrm{Sat}}}$ can be computed as
\begin{equation}\label{SINR_i}
	\begin{aligned}
		&\mathcal{L}_{I_2}(s_2)\\
		&=\exp \bigg(-\pi \lambda_2 \frac{R_s+R_e}{R_e} \bigg(\frac{s_2P_2G_2}{m_2}\bigg)\int_{(\frac{s_2P_2G_2}{m_2})^{-1}}^{(\frac{s_2P_2G_2}{m_2})^{-1}(\frac{R_{\max}}{R_{\min}})^2} \\
		&1-\frac{1}{(1+y^{-2})^{m_2}}  dy \bigg),
	\end{aligned}
\end{equation}
where $m_2=\frac{(K_2+1)^2}{2K_2+1}$ is  a given constant of $K_2$. Furthermore, the expression for the coverage probability in (\ref{P1cov3}) is accurate and sufficiently general across all relevant system parameters, but the impacts of relevant parameters to coverage probability is still not refined enough. In detail, the difficulty in deriving more elaborate expressions lies in calculating the Laplace transform of interference, which also requires integration. To get more clear expressions, we consider two special cases in this paper. In the first case that there are many obstacles on the sea, the direct path will occupy less of the total propagation path, so the Rician factor $K$ will also be small. Therefore, we assume that $K=2.414$ dB and evaluate $s_1=r^4\gamma_1$ and $s_2=r^2\gamma_2$, we obtain the lower bound of $\mathcal{L}_{I_1}(s_1)$ as
\begin{equation}
	\begin{aligned}
		\mathcal{L}_{I_1}^{l}(s_1)=\exp \bigg(-\pi \lambda_1 (\frac{s_1P_1G_1}{m_1})^{\frac{1}{2}}\eta_1^l(\frac{2\gamma_1}{P_1G_1}) \bigg),
	\end{aligned}
\end{equation}
where $\eta_1^l(x)= \frac{1}{2} \bigg(R_0^2+\frac{R_0^6}{x^2+R_0^4}\bigg)+3x\arctan\bigg(\frac{R_0^2}{x}\bigg)$. Similarly, we obtain
\begin{equation}
	\begin{aligned}
		\mathcal{L}_{I_2}^{l}(s_2)=\exp \bigg(-\pi \lambda_2 \frac{R_s+R_e}{R_e} \bigg(\frac{s_2P_2G_2}{m_2}\bigg)\eta_2^l(\frac{2\gamma_2}{P_2G_2}) \bigg),
	\end{aligned}
\end{equation}
where $\eta_2^l(x)=x\ln\bigg(\frac{x+\frac{2(2R_e+R_s)}{R_e}}{2+x}\bigg)+\frac{x^2}{2(\frac{2(2R_e+R_s)}{R_e}+x)}-\frac{x^2}{2(2+x)}$. In the second that there are few obstacles on the sea, the direct path occupies the main component of the total propagation path, so the Rician factor $K$ will also be larger. We assume that $K=6.464$ dB, then the upper bound of $\mathcal{L}_{I_1}(s_1)$ can be obtained as
\begin{equation}\label{SINR_i}
	\begin{aligned}
		\mathcal{L}_{I_1}^{h}(s_1)=\exp \bigg(-\pi \lambda_1 \bigg(\frac{s_1P_1G_1}{m_1}\bigg)^{\frac{1}{2}}\eta_1^h(\frac{4\gamma_1}{P_1G_1}) \bigg),
	\end{aligned}
\end{equation}
where $\eta_1^h(x)=\frac{R_0^{14}+14R_0^{10}(x^2+R_0^4)^2+35R_0^6(x^2+R_0^4)^2}{48(x^2+R_0^4)^3}-\frac{19}{16}R_0^2+\frac{35x}{18}\arctan\bigg(\frac{R_0^2}{x}\bigg)$. Similarly, we obtain
\begin{equation}\label{SINR_i}
	\begin{aligned}
		\mathcal{L}_{I_2}^{h}(s_2)=\exp \bigg(-\pi \lambda_2 \frac{R_s+R_e}{R_e} \bigg(\frac{s_2P_2G_2}{m_2}\bigg)\eta_2^h(\frac{4\gamma_2}{P_2G_2}) \bigg),
	\end{aligned}
\end{equation}
where $\eta_2^h(x)=\frac{18x^2(x+\frac{4(2R_e+R_s)}{R_e})^2-6x^3(x+\frac{4(2R_e+R_s)}{R_e})+x^4}{12(x+\frac{4(2R_e+R_s)}{R_e})^3}+x\text{ln}\bigg(\frac{x+\frac{4(2R_e+R_s)}{R_e}}{4+x}\bigg)-\frac{18x^2(x+4)^2-6x^3(x+4)+x^4}{12(x+4)^3}$. Then, we derive the coverage probability' closed-form formula of the near-shore users, which is given by
\begin{equation}
	\begin{aligned}
		&\mathcal{P}_n=\big[1-\exp(-\lambda_1\pi R_0^2)\big]2\pi \lambda_1 \exp(\lambda_1\pi R_0^2)\\
		& \bigg(\sum_{n=1}^{N}a_n^* e^{-\frac{b_n^*\gamma_1\sigma_1^2}{\zeta G_1P_1}} \times \frac{1-\exp(\lambda_1\pi R_0^2 \psi_1^l)}{\psi_1^l}\bigg)
		\label{Pn}
	\end{aligned}
\end{equation}
where $\psi_1^l=1+\eta_1^l+\eta_2^l$ represents the conditional Laplace transformation of the cumulative interference power, which is approximated based on the $K=2.414$ dB. Similarly, we can get the coverage probability's closed-form formula of the off-shore users, which is given at the top of next page as (\ref{Po}), where $\psi_2^l=1+\eta_2^l$ is the conditional Laplace transformation of the interference, which is approximated based on $K=2.414$ dB.

To get more clear insights, we perform asymptotic analysis of several parameters in the formula (\ref{Pn}) and (\ref{Po}).

1) High transmit power: Due to high transmit power, we have $\frac{P_1}{1+\sigma_{1}^2}\rightarrow \infty$ and $\frac{P_2}{1+\sigma_{2}^2}\rightarrow \infty$ \cite{K}. To facilitate the subsequent derivation and analysis, we recalculate the PDF and CCDF of the received signal-to-noise ratio at maritime users as

\begin{figure*}[ht]
	\begin{equation}
		\mathcal{P}_o=[1-e^{-2\lambda_2 \pi R_e(R_s+R_e)}]2\pi \lambda_2 \frac{(R_s+R_e)e^{\lambda_2\pi (R_e+R_e)^2}}{R_e(e^{2\lambda_2\pi R_e(R_s+R_e)}-1)} \bigg(\sum_{n=1}^{N}a_n^* e^{-\frac{b_n^*\gamma_2\sigma_2^2}{\zeta G_2P_2}} *\frac{e^{-\lambda_2\pi \frac{R_s+R_e}{R_e}\psi_2^lR_{\min}^2}-e^{-\lambda_2\pi \frac{R_s+R_e}{R_e}\psi_2^lR_{\max}^2}}{1+\psi_2^l}\bigg)
		\label{Po}
	\end{equation}
		\centering
	\rule[-10pt]{18cm}{0.07em}
\end{figure*}

\begin{equation}\label{f1}
	\begin{aligned}
		f_{\Omega}(x)=\frac{(1+K)}{\Omega}&J_0\bigg(2\sqrt{\frac{xK(1+K)}{\Omega}}\bigg)\\
		&\times \exp\bigg(-K-\frac{x(1+K)}{\Omega}\bigg),
	\end{aligned}
\end{equation}
and
\begin{equation}\label{F1}
	\begin{aligned}
		F_{\Omega}(x)=1-Q_1\bigg(\sqrt{2K},\sqrt{\frac{2(1+K)x}{\Omega}}\bigg),
	\end{aligned}
\end{equation}
where $\Omega=\frac{\overline{\lambda} P_i}{1+\sigma_0^2}$, $P_i$ is the transmit power of LEO satellite or TBS and $\overline{\lambda}$ denotes the variance of the signal. Thus, by adopting series representations for (\ref{f1}) and (\ref{F1}), we obtain
\begin{equation}
	\begin{aligned}
		\exp\bigg(-\frac{(1+K)x}{\Omega}\bigg) \approx 1- \frac{(1+K)x}{\Omega},
	\end{aligned}
\end{equation}

\begin{equation}
	\begin{aligned}
		J_0\bigg(2\sqrt{\frac{(1+K)K}{\Omega}}\bigg) \approx 1+\frac{(1+K)K}{\Omega}.
	\end{aligned}
\end{equation}

Thus, the PDF and CCDF of small scale fading of channel in (\ref{channel1}) and (\ref{channel2}) can be further simplified as
\begin{equation}\label{KPDF}
	\begin{aligned}
		\widetilde{f}_{\Omega}(x)=\frac{e^{-K}(1+K)}{\Omega}\bigg(1-\frac{x^2(1+K)}{\Omega}\bigg)\bigg(1+\frac{x^2(1+K)K}{\Omega}\bigg),
	\end{aligned}
\end{equation}
and
\begin{equation}\label{KCCDF}
	\begin{aligned}
		&\widetilde{F}_{\Omega}(x)=\\
		&\frac{e^{-K}(1+K)}{\Omega}\bigg(x^2-\frac{x^6(1+K)^2K}{6\Omega^2}+\frac{x^4(1+K)(1-K)}{4\Omega}\bigg).
	\end{aligned}
\end{equation}
Then, substituting (\ref{KPDF}) and (\ref{KCCDF}) into the formulas, we recalculate the asymptotic analysis of coverage probability for near-shore users as
\begin{equation}\label{SINR_i}
	\begin{aligned}
		&\mathcal{P}_n^{\infty}=[1-\exp(-\lambda_1\pi R_0^2)]2\pi \lambda_1 \exp(\lambda_1\pi R_0^2) \\
		&\bigg(\sum_{n=1}^{N}a_n^* e^{-\frac{b_n^*\gamma_1\sigma_1^2}{\zeta G_1P_1}}\bigg) *(A_1+A_2+A_3+A_4),
	\end{aligned}
\end{equation}
where $A_1=\frac{1-(1-a_1R_0^2-a_2R_0^4-a_3R_0^6)e^{-\lambda_1\pi R_0^2\eta_1^l}}{\eta_1^l}$, $A_2=\frac{(a_1+2a_2R_0^2+3a_3R_0^4)e^{-\lambda_1\pi R_0^2\eta_1^l}-a_1}{(\eta_1^l)^2}$, $A_3=\frac{(2a_2+6a_3R_0^2)e^{-\lambda_1\pi R_0^2\eta_1^l}-2a_2}{(\eta_1^l)^3}$, $A_4=\frac{6a_3e^{-\lambda_1\pi R_0^2\eta_1^l}-6a_3}{(\eta_1^l)^4}$, with $a_1=\frac{(1+K)e^{-K}}{\Omega}$, and $a_2=\frac{(1-K)(1+K)e^{-K}}{4\Omega^2}$ and $a_3=-\frac{K(1+K)^3e^{-K}}{6\Omega^3}$. Similarly, we can conduct asymptotic analysis of coverage probability at high transmit power for off-shore users, which is given as
\begin{equation}\label{SINR_i}
	\begin{aligned}
		\mathcal{P}_o^{\infty}=&\big[1-e^{-2\lambda_2 \pi R_e(R_s+R_e)}\big]2\pi \zeta \frac{(R_s+R_e)e^{\lambda_2\pi (R_e+R_e)^2}}{R_e(e^{2\lambda_2\pi R_e(R_s+R_e)}-1)} \\
		&\bigg(\sum_{n=1}^{N}a_n^* e^{-\frac{b_n^*\gamma_2\sigma_2^2}{\zeta G_2P_2}} \bigg) (B_1+B_2+B_3+B_4),
	\end{aligned}
\end{equation}
where $B_1=\frac{(1-a_1R_{min}^2-a_2R_{\min}^4-a_3R_{\min}^6)e^{\lambda_2\pi \frac{R_e+R_s}{R_e}\psi_2^lR_{\min}^2}}{\psi_2^l}-\frac{(1-a_1R_{\max}^2-a_2R_{\max}^4-a_3R_{\max}^6)e^{\lambda_2\pi \frac{R_e+R_s}{R_e}\psi_2^lR_{\max}^2}}{\psi_2^l}$, $B_2=\frac{(a_1+2a_2R_{\min}^2+3a_3R_{\min}^4)e^{\lambda_2\pi \frac{R_e+R_s}{R_e}\psi_2^lR_{\min}^2}}{\psi_2^l}-\frac{(a_1+2a_2R_{\max}^2+3a_3R_{\max}^4)e^{\lambda_2\pi \frac{R_e+R_s}{R_e}\psi_2^lR_{\max}^2}}{\psi_2^l}$, $B_3=\frac{(2a_2+6a_3R_{\min}^2)e^{\lambda_2\pi \frac{R_e+R_s}{R_e}\psi_2^lR_{\min}^2}}{\psi_2^l}-\frac{(2a_2+6a_3R_{\max}^2)e^{\lambda_2\pi \frac{R_e+R_s}{R_e}\psi_2^lR_{\max}^2}}{\psi_2^l}$, $B_4=\frac{6a_3e^{\lambda_2\pi \frac{R_e+R_s}{R_e}\psi_2^lR_{\min}^2}}{\psi_2^l}-\frac{6a_3e^{\lambda_2\pi \frac{R_e+R_s}{R_e}\psi_2^lR_{\max}^2}}{\psi_2^l}$, respectively. As mentioned above, it indicates that in high-density maritime areas, increasing power can improve performance of system within a certain range, but it will also increase interference and lead to performance degradation. Thus, it is necessary to balance the relationship between transmit power and interference.

2) $K \rightarrow 0$: When the direct path completely disappears, the communication between the transmitting base station and maritime users can only be communicated through other paths such as scattering refraction with the Rician factor $K \rightarrow 0$. Then the coverage probability can be approximately computed as
\begin{equation}\label{K01}
	\begin{aligned}
		\eta_1^{K \rightarrow 0}(x)=\sqrt{x}\bigg[\arctan\bigg(\frac{2R_0}{\sqrt{x}}\bigg)-\arctan\bigg(\frac{1}{\sqrt{x}}\bigg)\bigg],
	\end{aligned}
\end{equation}

\begin{equation}\label{K02}
	\begin{aligned}
		\eta_2^{K \rightarrow 0}(x)=x\ln\bigg(\frac{2R_e+R_s+x}{(R_e+R_e)x}\bigg).
	\end{aligned}
\end{equation}

3) $K\rightarrow \infty$: When the sea is relatively calm, the wave fluctuation is small, and there are fewer ships on the sea at this time, the channel between the TBS and maritime users is only through the direct path, so it can be degraded to the Gaussian channel with the Rician factor $K\rightarrow \infty$. Therefore, the coverage probability can be computed as
\begin{equation}\label{Kwuqiong1}
	\begin{aligned}
		\eta_1^{K\rightarrow \infty}(x)=e^{R_0^2x}-1,
	\end{aligned}
\end{equation}
\begin{equation}\label{Kwuqiong2}
	\begin{aligned}
		\eta_2^{K\rightarrow \infty}(x)=e^{\frac{(2R_e+R_s)x}{R_e}}-1.
	\end{aligned}
\end{equation}
By bringing (\ref{K01}), (\ref{K02}), (\ref{Kwuqiong1}) and (\ref{Kwuqiong2}) into (\ref{Pn}) and (\ref{Po}), we can obtain the asymptotic expressions of coverage probability $\mathcal{P}_n^{K \rightarrow 0}$, $\mathcal{P}_o^{K \rightarrow 0}$, $\mathcal{P}_n^{K \rightarrow \infty}$ and $\mathcal{P}_o^{K \rightarrow \infty}$ with $\psi_1^{0}=1+\eta_1^{K \rightarrow 0}+\eta_2^{K \rightarrow 0}$, $\psi_2^{0}=1+\eta_2^{K \rightarrow 0}$, $\psi_1^{\infty}=1+\eta_1^{K \rightarrow \infty}+\eta_2^{K \rightarrow \infty}$ and $\psi_2^{\infty}=1+\eta_2^{K \rightarrow \infty}$, respectively. Therefore, it makes sense to optimize the directivity of the antenna and beamforming in the off-shore maritime scenarios that the Rician factor K is high.

4) A large number of antennas: When TBS and LEO satellite are equipped with a relatively large-scale antennas, it makes communication channel hardening \cite{H1}. The eigenvalue distribution of the channel matrix tends to be concentrated, the number of channel conditions decreases, and the channel response becomes more stable and predictable. Specifically, the small-scale fading attenuates rapidly. Therefore, we use their means to replace the small-scale fading as \cite{H2}, \cite{H3}
\begin{equation}\label{junzhi}
	\begin{aligned}
		|h(K)|^2 \approx \mathbb{E}[|h(K)|^2]=\sum_{n=1}^{N}\frac{a_n^*}{b_n^*}.
	\end{aligned}
\end{equation}
In such a case, we can approximate the SINR by taking (\ref{junzhi}) into (\ref{SINR1}) and (\ref{SINR2}), which is given as
\begin{equation}
	\begin{aligned}
		\Upsilon_k^{1}=\frac{L_1r_k^{-4}}{\sum_{i\neq k}L_1r_i^{-4}+\sum_{m=1}L_2d_m^{-2}+\sigma_{1}^2},
	\end{aligned}
\end{equation}

\begin{equation}
	\begin{aligned}
		\Upsilon_j^{2}=\frac{L_2d_j^{-2}}{\sum_{n\neq j}L_2d_n^{-2}+\sigma_{2}^2},
	\end{aligned}
\end{equation}
where $L_1=(\frac{\zeta}{2\pi})^2G_1P_1\sum_{n=1}^{N}{a_n^*}/{b_n^*}$ and $L_2=g_r(\frac{\zeta}{2\pi})^2G_2P_2\sum_{n=1}^{N}{a_n^*}/{b_n^*}$. Then conditioning on $\Phi(\mathcal{A})>0$, we can compute the conditioned coverage probability as
\begin{equation}
	\begin{aligned}
	&\mathbb{P}\big(\Upsilon_k^1>\gamma_1|\Phi(\mathcal{A})>0\big)\\
	&{=}\mathbb{P}\bigg[1>\frac{\gamma_1(\sum_{i\neq k}L_1r_i^{-4}+\sum_{m=1}L_2d_m^{-2}+\sigma_{1}^2)}{L_1r_k^{-4}}\bigg]\\
	&\overset{(a)}{\approx}\mathbb{P}\bigg[\mu>\frac{\gamma_1(\sum_{i\neq k}L_1r_i^{-4}+\sum_{m=1}L_2d_m^{-2}+\sigma_{1}^2)}{L_1r_k^{-4}}\bigg]\\
	&\overset{(b)}{\approx} \sum_{q=1}^{Q}\omega_q\mathbb{E}\bigg[\exp\bigg(-\frac{\tau\gamma_1\sum_{i\neq k}L_1r_i^{-4}}{L_1r_k^{-4}}\\
	&\quad\quad\quad\quad\quad\quad    -\frac{\tau\gamma_1(\sum_{m=1}L_2d_m^{-2}+\sigma_{1}^2)}{L_1r_k^{-4}}\bigg)\bigg]\\
	&\overset{(c)}{\approx} \sum_{q=1}^{Q}\omega_q \exp \bigg\{-2\pi \lambda_1 \int_{x \in \mathcal{A}_t}\int_{y \in \mathcal{A}_r}  \bigg[1-\exp\bigg(\\
	&-\frac{\eta\gamma_1(\sum_{i\neq k}L_1x^{-4}+\sum_{m=1}L_2y^{-2}+\sigma_{1}^2)}{L_1r_k^{-4}}\bigg)\bigg]xydxdy\bigg\}\\
	&=\sum_{q=1}^{Q}\omega_q \exp\bigg[-\frac{\pi}{2}\lambda_1 (q\tau \gamma_1 )^{\frac{1}{2}} \beta_1r\\
	&\quad\quad\quad\quad\quad\quad    -\pi \lambda_2 (q\tau \gamma_1 )\beta_2r-(q\tau \gamma_1 )\sigma_1^2r^2      \bigg],
	\end{aligned}
\end{equation}
where $\omega_q=\binom{Q}{q}(-1)^{q+1} $, $\tau=Q(Q!)^{-1/Q}$, $\beta_1=\Gamma(-\frac{1}{2},q\tau \gamma_1)-\Gamma(-\frac{1}{2})$ and $\beta_2=\Gamma(-1,q\tau \gamma_1)-\Gamma(-1)$. Notice that in (a) we approximate the constant number one by a variable with shape parameter $Q$ and the unit mean, where $\tau$ converges to one when $Q$ tends to infinity. (b) follows from Alzer's inequality \cite{Alzer} and (c) comes from (\ref{M1}) and (\ref{M2}). Thus, we can get the coverage probability of near-shore users when channel hardening occurs as
\begin{equation}
 	\begin{aligned}
      &\mathcal{P}_n^h \approx \big[1-\exp(-\lambda_1\pi R_0^2)\big] \sum_{q=}^{Q} \omega_q \int_{0}^{R_0} \\ &\exp\bigg[-\frac{\pi}{2}\lambda_1 (q\tau \gamma_1 )^{\frac{1}{2}} \beta_1r-\pi \lambda_2 (q\tau \gamma_1 )\beta_2r-(q\tau \gamma_1 )\sigma_1^2r^2      \bigg]dr,
 	\end{aligned}
 \end{equation}

Similarly, we can get the coverage probability of off-shore users when channel hardening occurs as
\begin{equation}
	\begin{aligned}
		\mathcal{P}_o^h \approx &\big[1-e^{-2\lambda_2 \pi R_e(R_s+R_e)}\big] \sum_{q=}^{Q} \omega_q \\
		& \times \int_{R_{\min}}^{R_{\max}} \exp\bigg[-\pi \lambda_2 (q\tau \gamma_2 )\beta_2r-(q\tau \gamma_2 )\sigma_2^2r^2      \bigg]dr,
	\end{aligned}
\end{equation}

\subsection{Average Rate}
In this subsection, we focus on the mean data transmission rate experienced by the maritime users. We first present the fundamental rate theorem that characterizes the mean capacity for maritime communication terminals. In general, average rate is defined as $\mathbb{E}[\text{ln}(1+\text{SINR})]$, where the expectation is computed with respect to both the spatial PPP and the statistical characteristics of the fading channel. Furthermore, for a positive random variable $X$, we have $\mathbb{E}[X]=\int_{t>0}P(X>t)dt$. With this expression, the average rate of near-shore user $k$ can be calculated as
\begin{equation}
	\begin{aligned}
		&R_{1,k}=\mathbb{E}[\ln(1+\Upsilon_{1,k})]\\
		&=\int_{\mathcal{A}_r}f_{r_{\mathrm{TBS}|\Phi(\mathcal{A}_t)>0}}(r)\mathbb{E}\bigg\{\ln\bigg[1+\frac{\bigg(\frac{\zeta \sqrt{G_1P_1} }{2\pi d_k^2}|h(K_1)|\bigg)^2}{I_1+I_{\mathrm{Sat}}+\sigma_{1}^2}\bigg]\bigg\}dr\\
		&=\int_{\mathcal{A}_r}f_{r_{\mathrm{TBS}|\Phi(\mathcal{A}_t)>0}}(r)\int_{t>0} \\
		& \quad \quad\quad   P\bigg\{\ln\bigg[1+\frac{\bigg(\frac{\zeta \sqrt{G_1P_1} }{2\pi d_k^2}|h(K_1)|\bigg)^2}{I_1+I_{\mathrm{Sat}}+\sigma_{1}^2}\bigg]>t\bigg\}dtdr\\
		&=\int_{\mathcal{A}_r}e^{-\pi \lambda_1 r^2} \int_{t>0}e^{\sigma_1^2\gamma_1r^4(e^t-1)}\mathcal{L}_{I_1}(s_1)\mathcal{L}_{I_{2}}(s_2)dtdr,
	\end{aligned}
\end{equation}
by evaluating at $s_1=(e^t-1)r^4$ and $s_2=(e^t-1)r^2$, we obtain
\begin{equation}
	\begin{aligned}
	     \mathcal{L}_{I_1}(s_1)&=\exp\bigg(-2\pi\lambda_1\int_{0}^{R_0}(1-\frac{1}{1+(e^t-1)(\frac{r}{x})^4})xdx\bigg)\\
	     &=\exp\bigg(-r^2\alpha\arctan\big((e^t-1)^{-\frac{1}{2}}R_0\big)\bigg),
	\end{aligned}
\end{equation}
with $\alpha=\frac{\sqrt{2K+1}}{K+1}\sqrt{\zeta P_1G_1\gamma_1}\pi \lambda_1(e^t-1)^{\frac{1}{2}}$. Similarly, the Laplace form of $\mathcal{L}_{I_{\mathrm{Sat}}}$ can be expressed as
\begin{equation}
	\begin{aligned}
		\mathcal{L}_{I_2}(s_2)&=\exp\bigg(-2\pi\lambda_2\int_{R_{\min}}^{R_{\max}}(1-\frac{1}{1+(e^t-1)(\frac{r}{x})^2})xdx\bigg)\\
		&=\exp\bigg(-r^2\beta\ln\bigg(\frac{1+(e^t-1)R_{\max}}{1+(e^t-1)R_{\min}}\bigg)\bigg),
	\end{aligned}
\end{equation}
where $\beta=\frac{(2K+1)\zeta P_2G_2\gamma_1}{(K+1)^2}\lambda_1\pi \frac{R_s+R_e}{R_e}(e^t-1)$. In order to facilitate the calculation of average rate, we introduce a formulation which can be evaluated as
\begin{equation}\label{dingli}
	\begin{aligned}
		\int_{0}^{\infty}e^{-ax}e^{-bx^2}dx=\sqrt{\frac{\pi}{b}}\exp\bigg(\frac{a^2}{4b}\bigg)\mathcal{Q}\bigg(\frac{a}{\sqrt{2b}}\bigg),
	\end{aligned}
\end{equation}
where $\mathcal{Q}(x)=\frac{1}{\sqrt{2\pi}}\int_{x}^{\infty}\exp(-\frac{y^2}{x})dy$ denotes the standard Gaussian tail probability. By using (\ref{dingli}), the average rate can be further transformed as
\begin{equation}
	\begin{aligned}
		&R_{1,k}=\int_{0}^{\infty}\sqrt{\frac{\pi}{b_1(t)}}\exp\bigg(\frac{a_1(t)^2}{4b_1(t)}\bigg)\mathcal{Q}\bigg(\frac{a_1(t)}{\sqrt{2b_1(t)}}\bigg)dt,
	\end{aligned}
\end{equation}
where $a_1(t)=1+\alpha\arctan\big((e^t-1)^{-\frac{1}{2}}R_0\big)^{\frac{(K_1+1)^2}{2K_1+1}}+\beta\big(\ln(\frac{1+(e^t-1)R_{\max}}{1+(e^t-1)R_{\min}})\big)^{\frac{(K_1+1)^2}{2K_1+1}}$ and $b_1(t)=\frac{\sigma_1^2(e^t-1)}{(\pi \lambda_1)^2}$. Similarly, we can get the average rate of off-shore user $m$ as
\begin{equation}
	\begin{aligned}
		&R_{2,m}=\int_{0}^{\infty}\sqrt{\frac{\pi}{b_2(t)}}\exp\bigg(\frac{a_2(t)^2}{4b_2(t)}\bigg)\mathcal{Q}\bigg(\frac{a_2(t)}{\sqrt{2b_2(t)}}\bigg)dt,
	\end{aligned}
\end{equation}
where $a_2(t)=1+\beta\big(\ln(\frac{1+(e^t-1)R_{\max}}{1+(e^t-1)R_{\min}})\big)^{\frac{(K_2+1)^2}{2K_2+1}}$ and $b_2(t)=\frac{\sigma_2^2(e^t-1)}{(\pi \lambda_2)^2}$. We also perform asymptotic analysis on the average rate.

1) High transmit power: When the transmit power is very high, we have $\frac{P_1}{1+\sigma_1^2}\rightarrow \infty$ and $\frac{P_2}{1+\sigma_2^2}\rightarrow \infty$, then the average rate of near-shore user $k$ and off-shore user $m$ can be computed as
\begin{equation}
	\begin{aligned}
		&R_{1,k}^{\infty}=\\
		&\int_{0}^{\infty} \bigg\{1+\alpha\arctan\big((e^t-1)^{-\frac{1}{2}}R_0\big)+\\
		&\beta\bigg[\ln\big(1+(e^t-1)R_{\max}\big)-\ln\big(1+(e^t-1)R_{\min}\big)\bigg]\bigg\}^{-1}dt,
	\end{aligned}
\end{equation}

\begin{equation}
	\begin{aligned}
		&R_{2,m}^{\infty}=\int_{0}^{\infty} \frac{1}{1+\beta\bigg(\ln\big(\frac{1+(e^t-1)R_{\max}}{1+(e^t-1)R_{\min}}\big)\bigg)^{\frac{(K+1)^2}{2K+1}}}dt.
	\end{aligned}
\end{equation}

2) $K\rightarrow 0$: The maritime link only has NLOS component, and the channel degenerates to Rayleigh fading, $a_1^{K \rightarrow 0}(t)=1+\alpha\arctan((e^t-1)^{-\frac{1}{2}}R_0)+\beta(\ln(\frac{1+(e^t-1)R_{\max}}{1+(e^t-1)R_{\min}}))$ and $a_2^{K \rightarrow 0}(t)=1+\beta(\ln(\frac{1+(e^t-1)R_{\max}}{1+(e^t-1)R_{\min}}))$. Then the average rate of near-shore user $k$ and off-shore user $m$ can be expressed as
\begin{equation}
	\begin{aligned}
		&R_{1,k}^{K\rightarrow 0}=\int_{0}^{\infty}\sqrt{\frac{\pi}{b_1(t)}}\exp\bigg(\frac{a_1^{K\rightarrow 0}(t)^2}{4b_1(t)}\bigg)\mathcal{Q}\bigg(\frac{a_1^{K\rightarrow 0}(t)}{\sqrt{2b_1(t)}}\bigg)dt,
	\end{aligned}
\end{equation}

\begin{equation}
	\begin{aligned}
		&R_{2,m}^{K\rightarrow 0}=\int_{0}^{\infty}\sqrt{\frac{\pi}{b_2(t)}}\exp\bigg(\frac{a_2^{K\rightarrow 0}(t)^2}{4b_2(t)}\bigg)\mathcal{Q}\bigg(\frac{a_2^{K\rightarrow 0}(t)}{\sqrt{2b_2(t)}}\bigg)dt.
	\end{aligned}
\end{equation}
Thus, it is necessary to enhance the multipath diversity in the near-shore maritime scenarios that the Rician factor K is low.

3) $K\rightarrow \infty$: The channel will be degraded to the Gaussian channel with $a_1^{K \rightarrow \infty}(t)=1+\beta(\ln(\frac{1+(e^t-1)R_{\max}}{1+(e^t-1)R_{\min}}))$ and $a_2^{K \rightarrow \infty}=1$. Then the average rate can be obtained as
\begin{equation}
	\begin{aligned}
		&R_{1,k}^{K\rightarrow \infty}=\int_{0}^{\infty}\sqrt{\frac{\pi}{b_1(t)}}\exp\bigg(\frac{a_1^{K\rightarrow \infty}(t)^2}{4b_1(t)}\bigg)\mathcal{Q}\bigg(\frac{a_1^{K\rightarrow \infty}(t)}{\sqrt{2b_1(t)}}\bigg)dt,
	\end{aligned}
\end{equation}

\begin{equation}
	\begin{aligned}
		&R_{2,m}^{K\rightarrow \infty}=\int_{0}^{\infty}\sqrt{\frac{\pi}{b_2(t)}}\exp\bigg(\frac{a_2^{K\rightarrow \infty}(t)^2}{4b_2(t)}\bigg)\mathcal{Q}\bigg(\frac{a_2^{K\rightarrow \infty}(t)}{\sqrt{2b_2(t)}}\bigg)dt.
	\end{aligned}
\end{equation}

4) A large number of antennas: By replacing small-scale fading with their means, the average rate can be calculated directly as follows
\begin{equation}
	\begin{aligned}
		&R_{1,k}^{h}=\sum_{q=}^{Q} \omega_q\int_{0}^{\infty}\sqrt{\frac{\pi}{b_1(t)}}\exp\bigg(\frac{a_1^h(t)^2}{4b_1(t)}\bigg)\mathcal{Q}\bigg(\frac{a_1^h(t)}{\sqrt{2b_1(t)}}\bigg)dt,
	\end{aligned}
\end{equation}

\begin{equation}
	\begin{aligned}
		&R_{2,m}^{h}=\sum_{q=}^{Q} \omega_q\int_{0}^{\infty}\sqrt{\frac{\pi}{b_2(t)}}\exp\bigg(\frac{a_2^{h}(t)^2}{4b_2(t)}\bigg)\mathcal{Q}\bigg(\frac{a_2^{h}(t)}{\sqrt{2b_2(t)}}\bigg)dt,
	\end{aligned}
\end{equation}
where $a_1^h(t)=1+q\tau \beta_1(e^t-1)^{\frac{1}{2}}+q\tau\beta_2 (e^t-1) +q\tau (e^t-1) \sigma_1^2$  and $a_2^h(t)=1+q\tau\beta_2 (e^t-1) +q\tau (e^t-1) \sigma_2^2$. We can find that we cannot simply keep increasing the number of antennas when deploying base stations. It is necessary to balance the number of antennas, power consumption, system performance and the complexity of beam management.

With the above performance expressions, we can evaluate the performance of maritime hybrid network directly. Moreover, the efficiency of hybrid maritime communication network can be enhanced through systematic optimization of critical system parameters.

\section{Coverage Probability Optimization}

In this section, we strive to maximize the coverage probability of near-shore users by optimizing system parameters. Specifically, the coverage probability of near-shore users is maximized by balancing the signal strength received by nearshore users and the interference brought by LEO satellites.

We choose the coverage probability of near-shore users as the optimization objective. By adjusting the transmission power of LEO satellite and TBS, we aim to maximize the coverage probability of near-shore users. Furthermore, maritime users should be ensured to meet the minimum communication rate requirements. Specifically, the optimization problem can be formulated as
\begin{subequations}
	\begin{eqnarray}
		\mathcal{Q}1:	\underset{P_1,P_2}{\mathop{\max}}\,\!\!\!\!\!\!\!&&\!\! \mathcal{P}_n \label{OP2obj}\\
		\textrm{s.t.}&&\!\!\!\!\!\!  \ln(1+\Upsilon_1)\geq r_1,\label{OP2st1}\\
		&&\!\!\!\!\!\!  \ln(1+\Upsilon_2)\geq r_2,\label{OP2st1}\\
		&&\!\!\!\!\!\! 0 \leq P_1\leq P_1^{\max},\label{OP2st2}\\
		&&\!\!\!\!\!\! 0 \leq P_2\leq P_2^{\max}, \label{OP2st3}
	\end{eqnarray}
\end{subequations}
where $r_1$ and $r_2$ are the minimum data rate requirements of near-shore and off-shore users, $P_1^{\max}$ and $P_2^{\max}$ denote the maximum transmission power of TBS and LEO satellite, respectively. Noticeably, since the objective function $\mathcal{P}_n$ is highly coupled to $P_1$ and $P_2$, it is not convex. Therefore, we first deal with the objective function. By applying the Taylor expansion and approximation of the aggregate interference, we can transform the objective function as follows
\begin{equation}
	\begin{aligned}
		&\mathcal{P}_n=\big[1-\exp(-\lambda_1\pi R_0^2)\big]2\pi \lambda_1 \exp(\lambda_1\pi R_0^2) \\
		&\bigg(\sum_{n=1}^{N}a_n^* e^{-\frac{b_n^*\gamma_1\sigma_1^2}{\zeta G_1P_1}} \times \frac{1-\exp(\lambda_1\pi R_0^2 \psi_1^l)}{\psi_1^l}\bigg)\\
		&\overset{(a)}{\approx}\big[1-\exp(-\lambda_1\pi R_0^2)\big]2\pi \lambda_1 \exp(\lambda_1\pi R_0^2)\\ &\bigg(\sum_{n=1}^{N}a_n^* e^{-\frac{b_n^*\gamma_1\sigma_1^2}{\zeta G_1P_1}} \times \frac{1-1-\lambda_1\pi R_0^2 \psi_1^l-\frac{(\lambda_1\pi R_0^2 \psi_1^l)^2}{2}}{\psi_1^l}\bigg)\\
		&=\big[1-\exp(-\lambda_1\pi R_0^2)\big]2\pi \lambda_1 \exp(\lambda_1\pi R_0^2) \\
		&\bigg(\sum_{n=1}^{N}a_n^* e^{-\frac{b_n^*\gamma_1\sigma_1^2}{\zeta G_1P_1}} \times \bigg(-\lambda_1\pi R_0^2-\frac{(\lambda_1\pi R_0^2)^2\psi_1^l}{2}\bigg)\bigg)\\
		&\overset{(b)}{\approx}\big[1-\exp(-\lambda_1\pi R_0^2)\big]2\pi \lambda_1 \exp(\lambda_1\pi R_0^2)\\ &\bigg(\sum_{n=1}^{N}a_n^* e^{-\frac{b_n^*\gamma_1\sigma_1^2}{\zeta G_1P_1}} \times (\eta_1-\eta_2)\bigg)\\
		&\overset{(c)}{\approx}c_1(e^{-c_2\frac{P_2}{P_1}}-e^{-c_3\frac{P_2}{P_1}})
		\label{Pn}
	\end{aligned}
\end{equation}
where (a) follows from the Taylor expansion of the series and in the (b) we approximate the aggregated interference as $\eta_1=e^{-\pi \lambda_1 R_0^2 \frac{P_2}{P_1}}$ and $\eta_2=e^{-\pi \lambda_1 \frac{(2R_e+R_s)P_2}{R_eP_1} }$, respectively, which are based on the assumption that $R_0$ ia large enough and the Rician factor $K = 2.414$ dB. Furthermore, in the (c) we introduce new variables $c_1=2\pi \lambda_1[e^{\lambda_1\pi R_0^2}-1]\sum_{n=1}^{N}a_n^* e^{-\frac{b_n^*\gamma_1\sigma_1^2}{\zeta G_1}}$, $c_2=\pi \lambda_1 R_0^2$ and  $c_3=\pi \lambda_1 \frac{2R_e+R_s}{R_e}$. And then we take the derivative of $P_n$ with respect to $z=\frac{P_2}{P_1}$, which is computed as

\begin{equation}
	\begin{aligned}
		\frac{\partial \mathcal{P}_n}{\partial z} = c_1(-c_2e^{-c_2z}+c_3e^{-c_3z}),
	\end{aligned}
\end{equation}
where we can find that $\mathcal{P}_n$ is unimodal function with respect to $z>0$. Therefore, $\mathcal{P}_n$ has a unique maximum at a certain power ratio. Nevertheless, the derived power allocation ratio may not satisfy the total transmission power requirements of stations and the required data rate for maritime users. Then we maximize the $\mathcal{P}_n$ by optimizing the incremental part of it, and transform this part according to the Dinkelbach's theorem. Then we make a simple transformation of the rate constraints, $P_1$ and $P_2$ are brought to one side of the inequality, where the new optimization problem can be obtained as follows

\begin{subequations}
	\begin{eqnarray}
		\mathcal{Q}2:	\underset{P_1,P_2}{\mathop{\max}}\,\!\!\!\!\!\!\!\!\!&&\! 2y\ln\bigg(\frac{c_3P_2}{c_2P_1}\bigg)-y^2(e^{c_3P_2}+e^{c_2P_1}-1) \label{OP3obj}\\
		\textrm{s.t.}&&\!\!  P_1\geq \sqrt{\frac{(e^{r_1}-1)(I_\mathrm{sat}+\sigma_1^2)}{(h_1r^{-4})^2-(e^{r_1}-1)I_\mathrm{TBS}}},\label{OP3st1}\\
		&&\!\!\!\!\!\!  P_2 \geq \sqrt{\frac{(e^{r_2}-1)\sigma_2^2}{(h_2r^{-2})^2-(e^{r_1}-1)I_\mathrm{sat}}},\label{OP3st2}\\
		&&\!\!\!\!\!\!  0 \leq P_1\leq P_1^{\max},\label{OP3st3}\\
		&&\!\!\!\!\!\!  0 \leq P_2\leq P_2^{\max}, \label{OP3st4}
	\end{eqnarray}
\end{subequations}
where the optimal $y$ can be found in closed form as $y^*=\frac{\ln(c_3P_2)-\ln(c_2P_1)}{(e^{c_3P_2}+e^{c_2P_1}-1)}$ when $P_1$ and $P_2$ are held fixed. By using CVX, the problem can be efficiently resolved. In conclusion, the computational procedure for solving the problem ($\mathcal{Q}2$) is systematically outlined in Algorithm 1.

\begin{algorithm}[h]
	\caption{Optimization for Coverage Probability of Near-Shore Users}
	\label{alg:AOA}
	\renewcommand{\algorithmicrequire}{\textbf{Input:}}
	\renewcommand{\algorithmicensure}{\textbf{Output:}}
	\begin{algorithmic}[1]
		\REQUIRE $\lambda_1$, $\lambda_2$, $G_1$, $G_2$, $R_0$, $R_{\min}$, $R_{\max}$, $r_1$, $r_2$, $\sigma^2_1$, $\sigma_2^2$, $\varepsilon_0$,  
		\ENSURE $P_1$, $P_2$    
		{\tiny }
		\STATE $\mathbf{Initialization:}$ Set transmit power $P_1^{(0)}$ and $P_2^{(0)}$, maximal iteration number $t_{\max}$, accuracy $\epsilon_0$
		\STATE $\mathbf{repeat}$
		\STATE \ \  Use CVX to address problem ($\mathcal{Q}2$), get $P_1^{(t)}$ and $P_2^{(t)}$
		\IF {$\bigg|2y\ln\bigg(\frac{c_3P_2^{(t)}}{c_2P_1^{(t)}}\bigg)-y^2(e^{c_3P_2^{(t)}}+e^{c_2P_1^{(t)}}-1)\bigg|>\epsilon_0$}
		\STATE Set $t=t+1$
		\ENDIF
		\STATE Update $y^{(t)}$ by $y^{(t+1)}=\frac{\ln\big(c_3P_2^{(t+1)}\big)-\ln\big(c_2P_1^{(t+1)}\big)}{\big(e^{c_3P_2^{(t+1)}}+e^{c_2P_1^{(t+1)}}-1\big)}$
		\STATE problem is solved or $\mathbf{until}$ $t=t_{\max}$
		\STATE get $P_1$ and $P_2$, and plug them into $\mathcal{P}_n$.
	\end{algorithmic}
\end{algorithm}

For ($\mathcal{Q}2$), each individual optimization variable is convex and the proposed algorithm can achieve convergence to stationary points for each variable through iterative computation, ultimately yielding the optimal solution to the optimization problem. Since $P_1$ and $P_2$ have a upper bound, it indicates that the proposed algorithm will ultimately converges after limited iterations. For computational complexity, it mainly comes from the constraints (\ref{OP3st1}) and (\ref{OP3st2}). By using the classical interior-point method (IPM) \cite{ym}, the computational complexity of the problem can be calculated as $\sqrt{2(M_1+1)}\cdot n\cdot \lbrace 2[(M_1+1)^3+1]+2n[(M_1+1)^2+1]+n^2\rbrace$, where $n=\mathcal{O}(M_1^2)$. Thus, the proposed algorithm is capable of deriving the near-optimal solutions within polynomial time complexity.

\section{Simulation Results}
In this section, we provide extensive simulation results to validate the derived performance metrics. Besides, we also evaluate the influence of key system parameters, including the altitude of LEO satellite, Rician factor, number of antennas and transmission power. The parameters of simulation are listed in Table \uppercase\expandafter{\romannumeral1}.

\begin{table}[h]
	\small
	\centering
	\caption{Main Simulation Parameters For Integrated Satellite-Terrestrial Communication Systems}\label{Simulation}
	\begin{tabular}{|c|c|}
		\hline
		Parameter & Value\\\hline
		Satellite orbit & LEO \\\hline
		Bandwidth  &  20 MHz \\\hline
		TBS antenna gain   &  25 dBi \\\hline
		Satellite antenna gain   &  50 dBi \\\hline
		Number of TBS antennas  &  8  \\\hline
		Number of LEO satellite antennas  &  8  \\\hline
		Carrier frequency &  160 MHz \\\hline
		Rain fading mean  & -2.6 dB  \\\hline
		Rain fading variance  & 1.63 dB  \\\hline
		Altitude of orbit      &  600 km  \\\hline
		Boltzmann's constant  &  1.38 $\times 10^{-23}$ J/m\\ \hline
		Variance of AWGN & -110 dBm\\ \hline
		Noise temperature   &  300 K \\\hline
		3dB angle & $0.4^{\circ}$ \\\hline
		TBS total transmit power   & 47 dBm \\ \hline
		Effective antennas height of TBS   & 50 m \\\hline
		Effective antennas height of users  & 10 m \\\hline
	\end{tabular}
\end{table}	

\begin{figure}[h]
	\centering
	\includegraphics[width=0.45\textwidth]{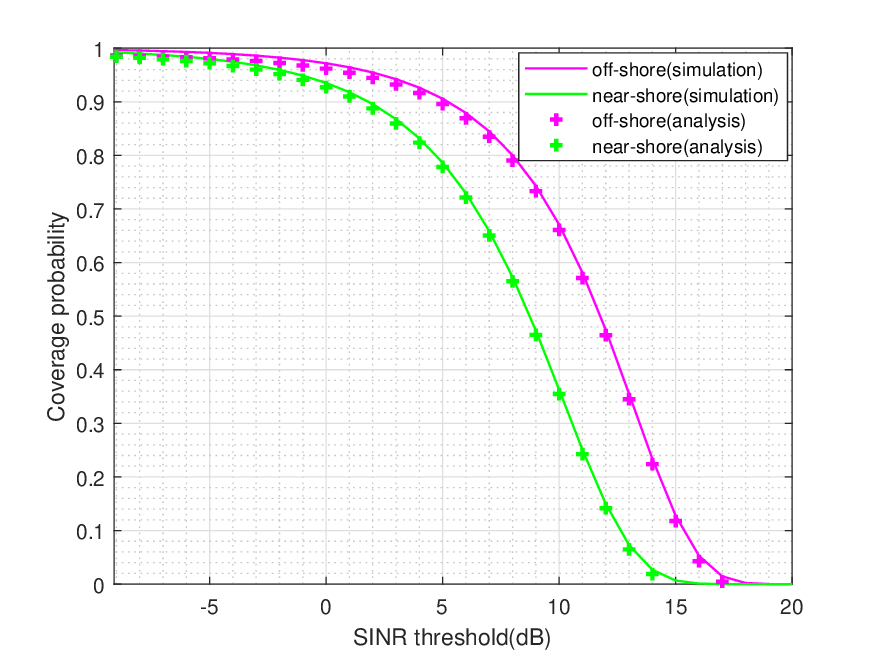} 
	\caption{Comparison of analysis and simulation of the coverage probability.}
	\label{yanzheng}
\end{figure}

\begin{figure}[h]
	\centering
	\includegraphics[width=0.45\textwidth]{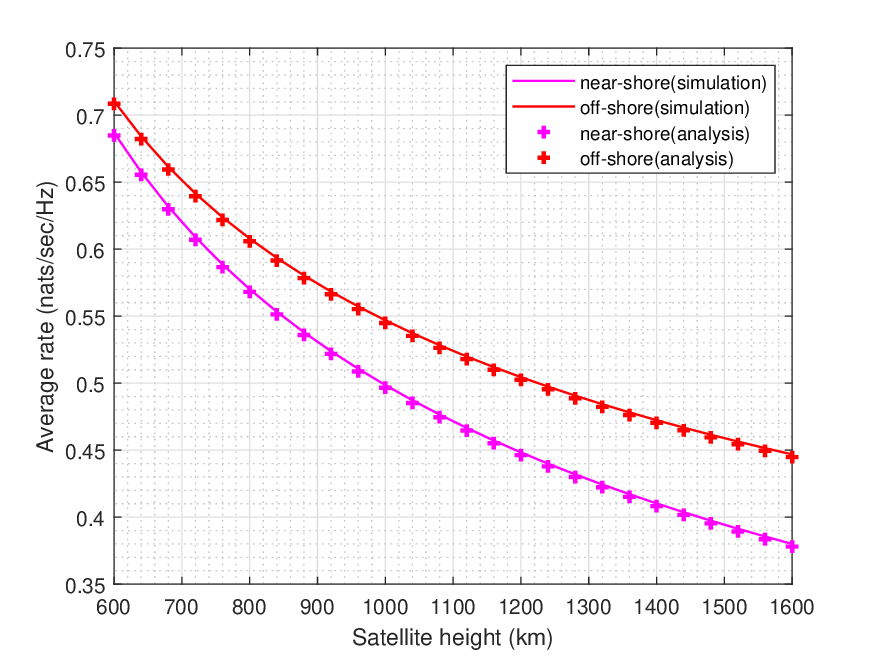} 
	\caption{Comparison of analysis and simulation of the average rate.}
	\label{Ryanzheng}
\end{figure}

Firstly, it is shown in Fig. $\ref{yanzheng}$ and Fig. $\ref{Ryanzheng}$ that the analytical expressions exactly match the numerical results for the average rate and coverage probability for near-shore and off-shore users, respectively. The validity of our analytical framework is substantiated through comprehensive simulation. Fig. $\ref{yanzheng}$ demonstrates that the coverage probability varies with the SINR threshold. Additionally, we can find that the coverage probability of near-shore users is consistently lower than that of off-shore users. This is due to the interference of near-shore users experience not only from nearby users but also from the LEO satellite. Thus, the SINR and data rate of near-shore users are always lower as shown in Fig. 3. Under the same SINR threshold, the coverage probability of near-shore users is smaller than that of off-shore users.

\begin{figure}[t]
	\centering
	\includegraphics[width=0.45\textwidth]{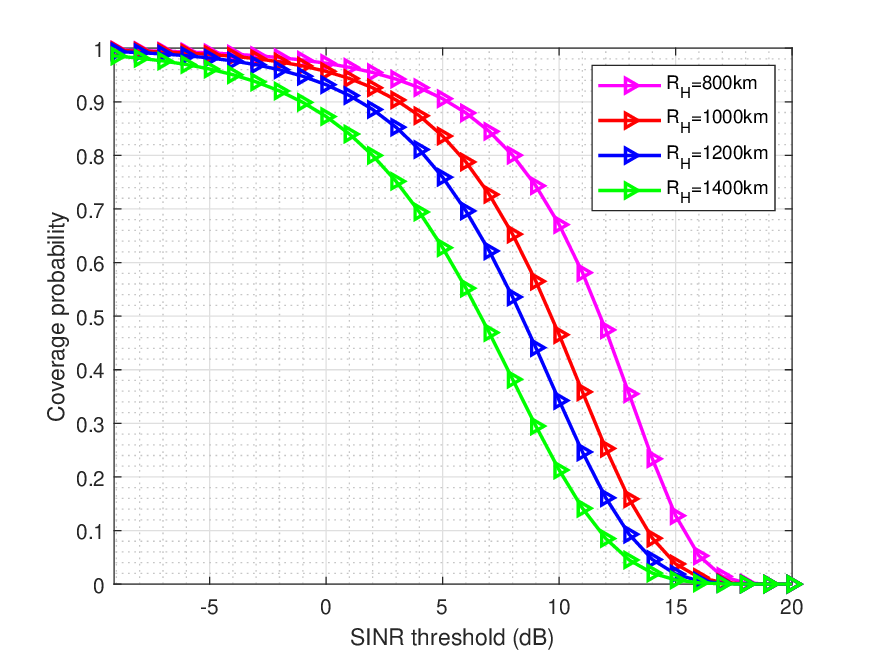} 
	\caption{Coverage probability versus SINR threshold with different LEO satellite altitudes of off-shore users.}
	\label{gaodu1}
\end{figure}

\begin{figure}[t]
	\centering
	\includegraphics[width=0.45\textwidth]{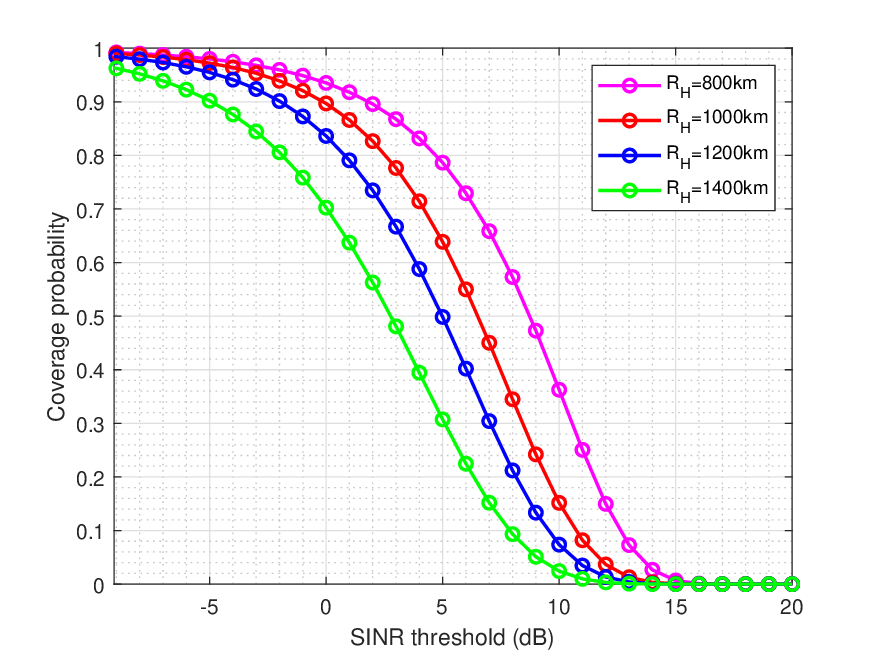} 
	\caption{Coverage probability versus SINR threshold with different LEO satellite altitudes of near-shore users.}
	\label{gaodu2}
\end{figure}

Moreover, Fig. $\ref{gaodu1}$ and Fig. $\ref{gaodu2}$ show the relationship between the coverage probability and the SINR threshold when the LEO satellite altitude is set to a series of altitudes. It is observed that as the LEO satellite altitude rises, the coverage probability decreases. This is because the increased distance between the LEO satellite and maritime users leads to greater path loss in the communication channel, which significantly reduces the quality of the space propagation link.

\begin{figure}[t]
	\centering
	\includegraphics[width=0.45\textwidth]{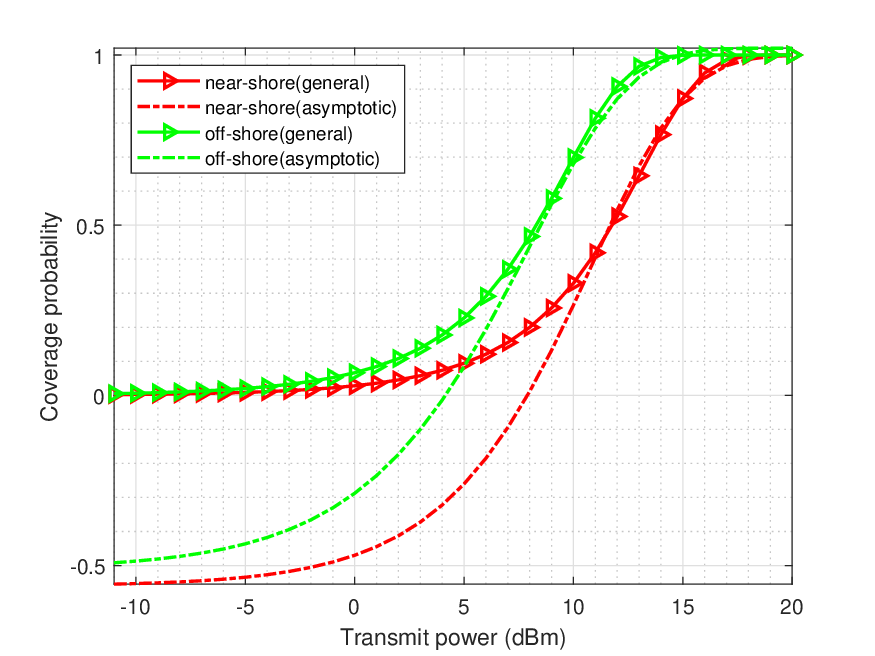} 
	\caption{Coverage probability versus transmit power with theorem and asymptotic analysis of near-shore and off-shore users.}
	\label{Pjianjin}
\end{figure}

\begin{figure}[t]
	\centering
	\includegraphics[width=0.45\textwidth]{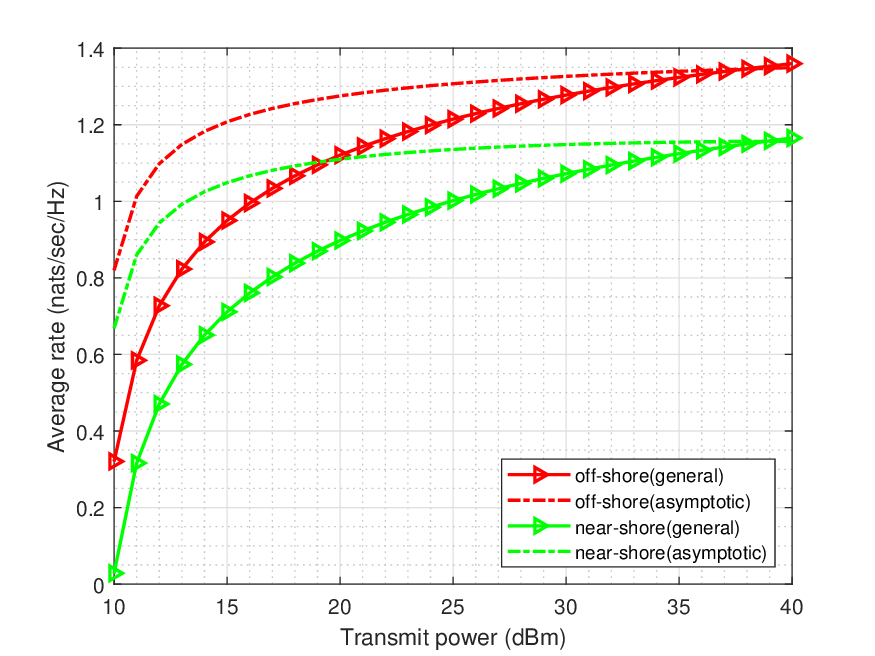} 
	\caption{Average rate versus transmit power with theorem and asymptotic analysis of near-shore and off-shore users.}
	\label{RPjianjin}
\end{figure}

\begin{figure}[t]
	\centering
	\includegraphics[width=0.45\textwidth]{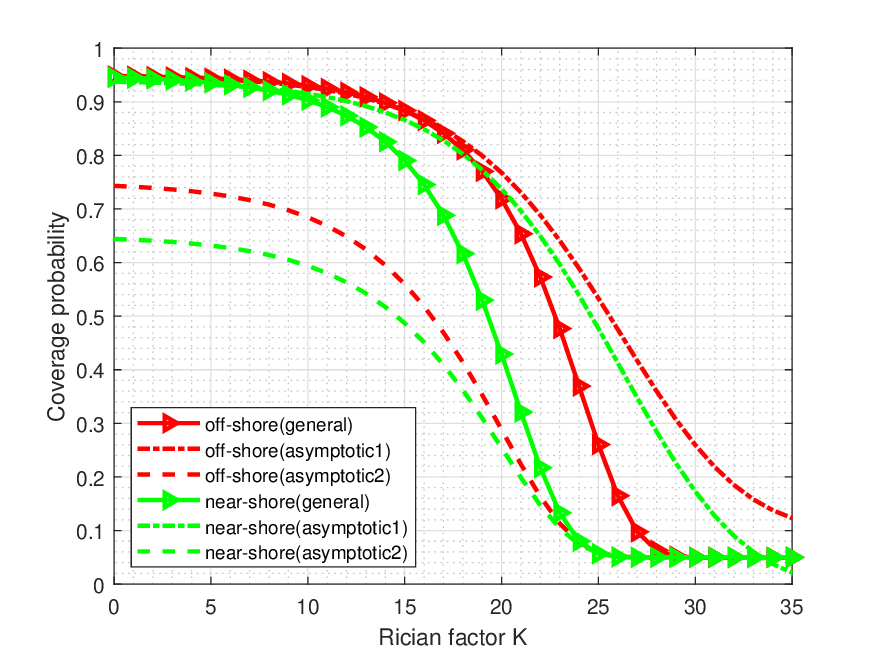} 
	\caption{Coverage probability versus Rician factor with theorem and asymptotic analysis of near-shore and off-shore users.}
	\label{Kjianjin}
\end{figure}

\begin{figure}[t]
	\centering
	\includegraphics[width=0.45\textwidth]{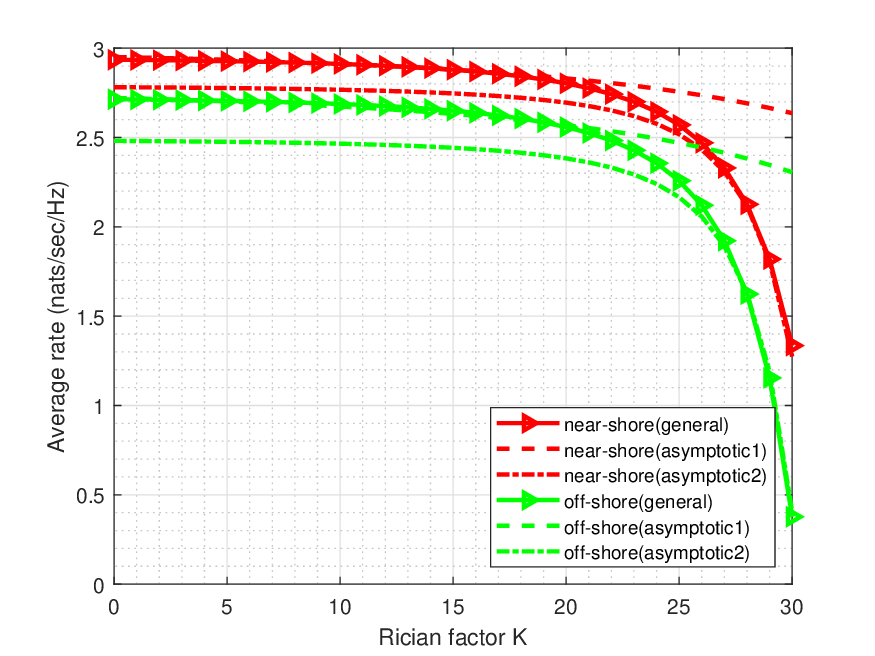} 
	\caption{Average rate versus Rician factor with theorem and asymptotic analysis of near-shore and off-shore users.}
	\label{RKjianjin}
\end{figure}

Next, we examine the effects of transmission power on the average rate and coverage probability. It is observed in Fig. $\ref{Pjianjin}$ and $\ref{RPjianjin}$ that as the transmission power increases, both the average rate and coverage probability improve progressively. Furthermore, the asymptotic results for the average rate and coverage probability match the simulation results accurately in the high transmission power regime. When transmit power exceeds a certain threshold, we can use the expression derived from the asymptotic analysis to approximate the theoretical formula for simplicity.

Then, we compare the relationship of the coverage probability and average rate with Rician factor $K$. From Fig. $\ref{Kjianjin}$ and Fig. $\ref{RKjianjin}$, it can be observed that as the $K$ reises, both the average rate and coverage probability gradually decrease. This is because, when wave fluctuations become large or obstacles increase, the LoS component between the TBS, LEO satellite, and maritime users is obstructed, leading to communication being solely reliant on NLoS paths such as refraction and scattering. However, the NLoS component inevitably weakens the signal strength. Besides, the asymptotic results for the average rate and coverage probability closely match the simulations in the $K\rightarrow 0$ and $K\rightarrow \infty$ regimes, respectively.

\begin{figure}[t]
	\centering
	\includegraphics[width=0.45\textwidth]{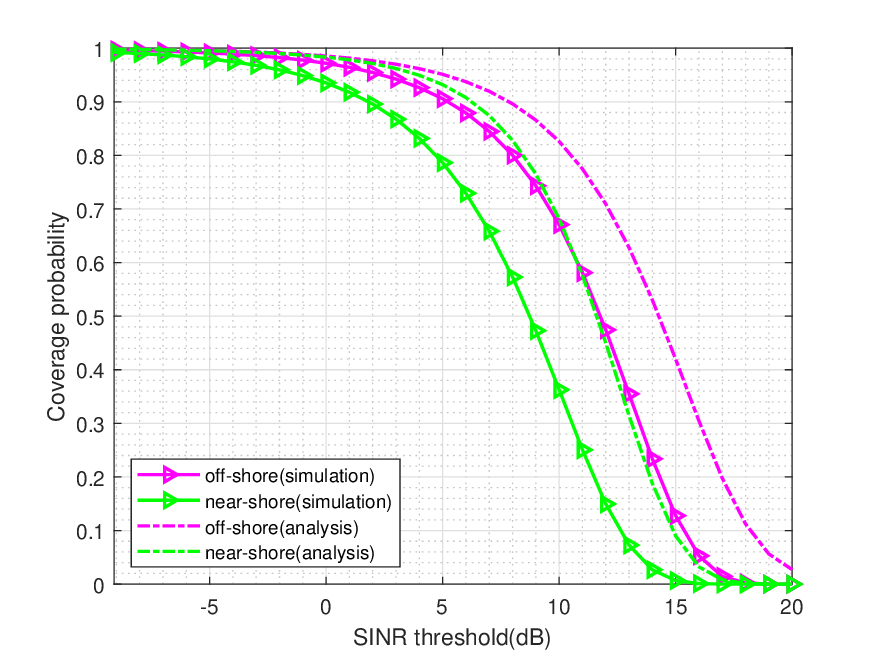} 
	\caption{Coverage probability versus SINR threshold with theorem and asymptotic analysis of near-shore and off-shore users.}
	\label{Hjianjin}
\end{figure}

\begin{figure}[t]
	\centering
	\includegraphics[width=0.45\textwidth]{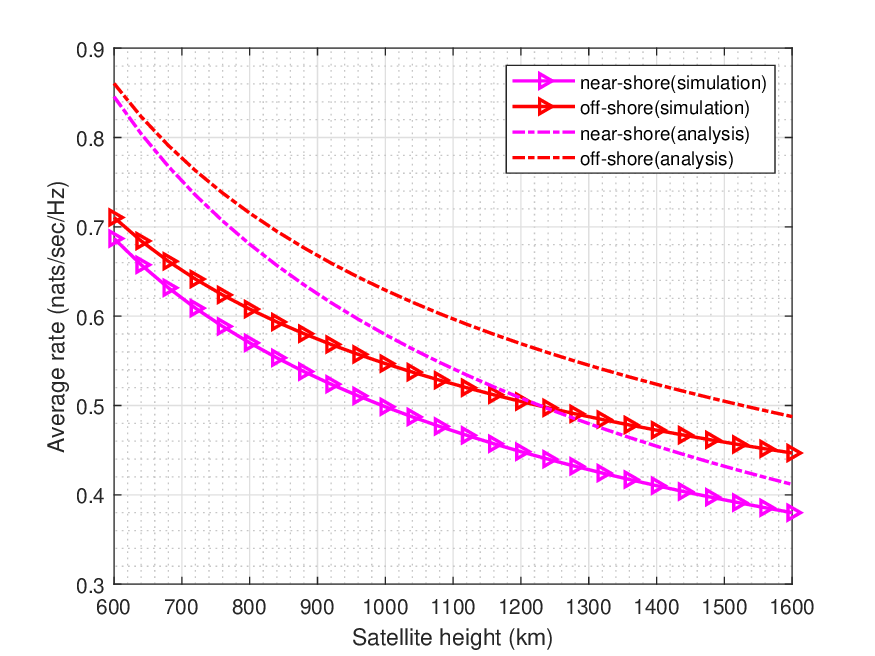} 
	\caption{Average rate versus SINR threshold with theorem and asymptotic analysis of near-shore and off-shore users.}
	\label{RHjianjin}
\end{figure}

Furthermore, we analyze the effect of a sharp increase in the number of antennas on the average rate and coverage probability. As shown in Figs. $\ref{Hjianjin}$ and $\ref{RHjianjin}$, both the transmission rate and coverage probability rise when channel hardening occurs. This is because that as the antenna array expands, the TBS and LEO satellite can use more free radicals, which can optimize the overall performance of the communication system.

\begin{figure}[t]
	\centering
	\includegraphics[width=0.45\textwidth]{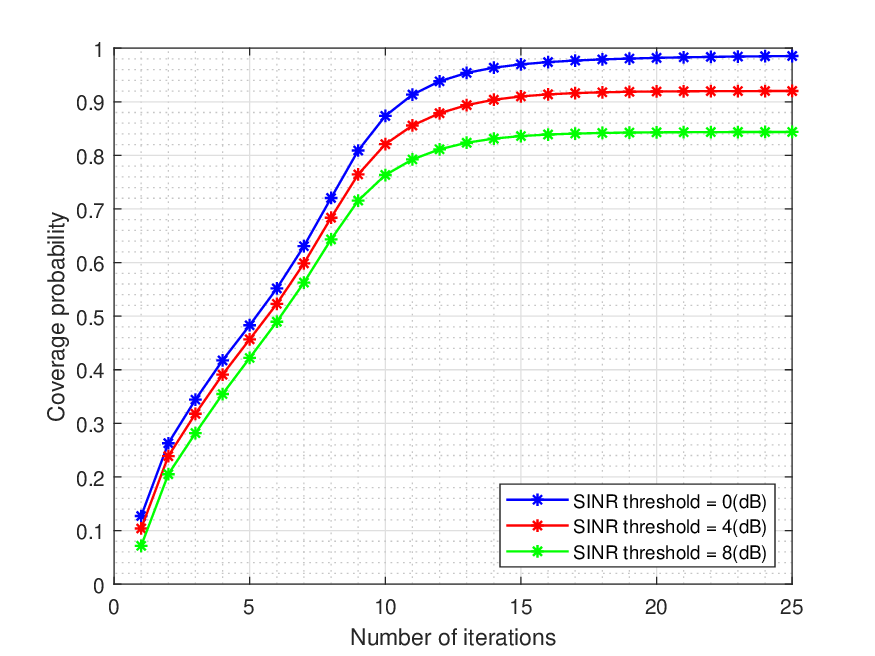} 
	\caption{Convergence performance of the proposed algorithm with different SINR thresholds.}
	\label{diedai}
\end{figure}

Then, we verify the effectiveness of the proposed algorithm. In Fig. $\ref{diedai}$, it illustrates the convergence of the coverage probability for SINR thresholds with 0 dB, 4 dB and 8 dB respectively. As shown in Fig. $\ref{diedai}$, the coverage probability gradually increases and achieves convergence to an equilibrium state in fewer than 25 iterative steps. The results demonstrate the efficient convergence characteristics of the proposed algorithm, and demonstrate that as the SINR threshold rises, the coverage probability decreases.

\begin{figure}[t]
	\centering
	\includegraphics[width=0.45\textwidth]{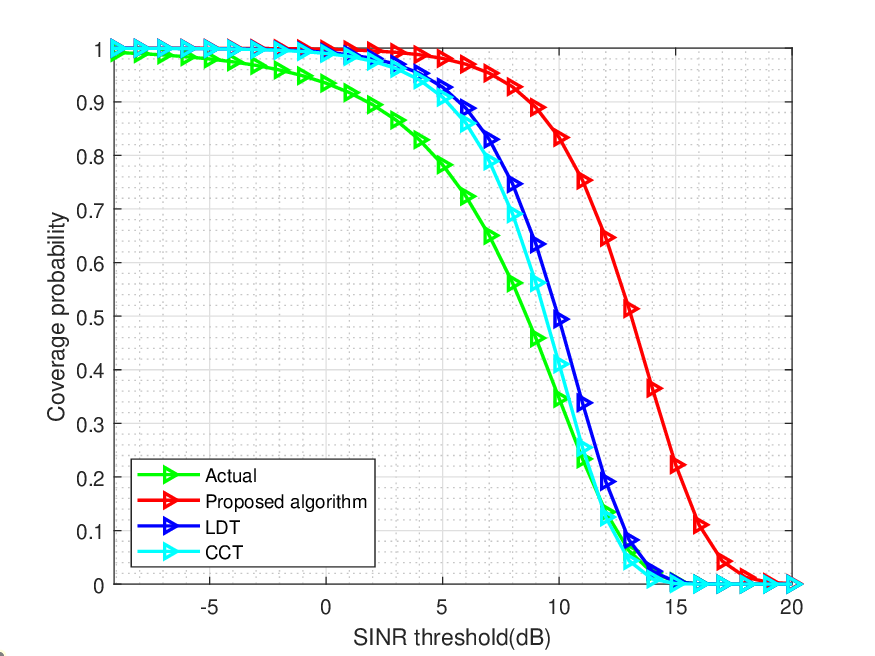} 
	\caption{Comparison of coverage probability with different values of SINR threshold between proposed algorithm and other algorithms.}
	\label{duibi}
\end{figure}

Finally, we compared the proposed algorithm with some baseline algorithms under the same conditions. The Lagrangian-Dual transform (LDT) \cite{DB1} solves the logarithmic function of the fraction by introducing Lagrangian duality. Another Charnes-Cooper transform (CCT) \cite{DB2} is the classical technique that is well for the single-ratio case. In Fig. $\ref{duibi}$, It can be observed that the proposed algorithm of optimized coverage probability is greatly enhanced compared to the derived coverage probability and the other algorithms.

\begin{figure}[t]
	\centering
	\includegraphics[width=0.45\textwidth]{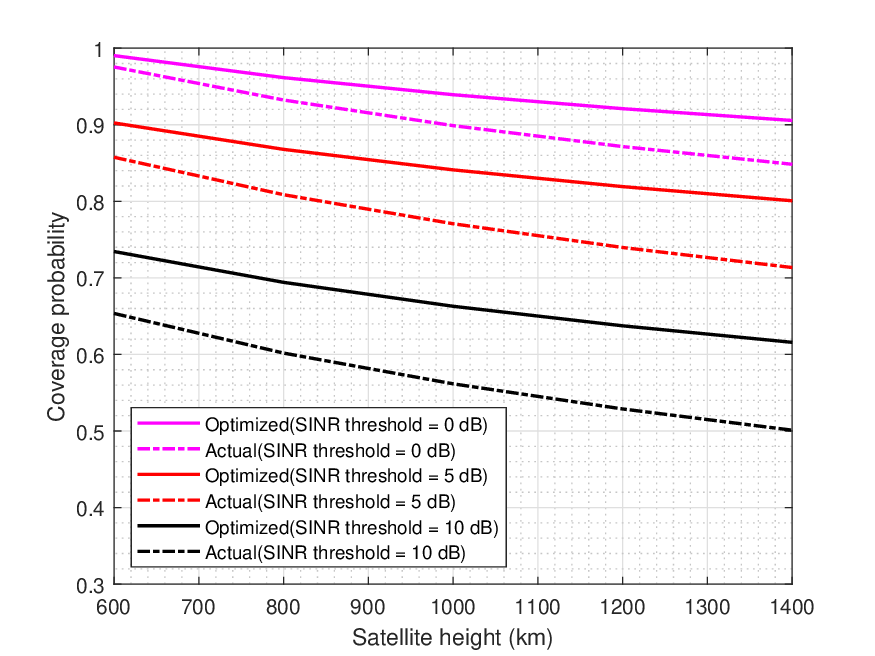} 
	\caption{Coverage probability of near-shore users versus satellite heights with different SINR thresholds.}
	\label{huifu}
\end{figure}

Besides, it is shown in Fig. \ref{huifu} that the coverage probability versus the satellite height with different SINR thresholds. It can be seen that the proposed algorithm has a great improvement in coverage probability compared with the actual theory under different SINR thresholds, thereby validating the effectiveness of the proposed algorithm. Additionally, this result has practical implications for the design of maritime communication networks, highlighting the necessity of optimizing coverage for near-shore users. It is essential to design the transmit power of TBS and LEO satellite. Proper coordination of the relationship between signal strength and interference can significantly improve both the communication quality and coverage probability for maritime users.

\section{Conclusion}
In this paper, we proposed an hybrid terrestrial satellite maritime network which can provide network services for both near-shore and off-shore users. Moreover, we derived performance analysis and asymptotic analysis on the average rate and coverage probability for both near-shore and off-shore users. Subsequently, we improve the coverage probability of near-shore users by jointly optimizing the transmission power of LEO satellite and TBS, while ensuring both off-shore and near-shore users can meet the minimum data requirements. Finally, the simulation results corroborate the derived formula’s validity, and the asymptotic analysis results can accurately match the simulation and analysis. Finally, simulation results show that the proposed algorithm can effectively enhance the coverage performance of near-shore users compared with theoretical analysis. In the future works, we will consider multi-satellite communication network where we will design and optimize constellations to deal with regulatory and deployment challenges.

\end{document}